\documentclass[structabstract]{aa}

\usepackage{graphicx}
\usepackage{amsmath}
\usepackage{polynom}
\usepackage{mdwlist}
\usepackage{url}
\usepackage{natbib,natbibspacing}
\usepackage{authblk}

\begin{document}

\let\oldthefootnote\thefootnote
\renewcommand{\thefootnote}{\fnsymbol{footnote}}
\let\thefootnote\oldthefootnote

\title{\texttt{MYRIAD}: A new N-body code for simulations of star clusters}
\author{Simos Konstantinidis\thanks{simos@astro.auth.gr}\inst{1,2} \and Kostas D. Kokkotas \inst{1,2}}
\institute{Department of Physics, Aristotle University of Thessaloniki, Thessaloniki 54124, Greece  \and  Theoretical Astrophysics, Eberhard-Karls University of T\"{u}bingen, T\"{u}bingen 72076, Germany}
\authorrunning{S. Konstantinidis \and K. D. Kokkotas}

\abstract
{}{We present a new C++ code for collisional N-body simulations of star clusters.} 
{The code uses the Hermite fourth-order scheme with block time steps, to advance the particles in time, while the forces and neighboring particles are computed using the GRAPE-6 board. Special treatment is used for close encounters and binary or multiple subsystems that form either dynamically or exist in the initial configuration. The structure of the code is modular and allows the appropriate treatment of more physical phenomena, such as stellar and binary evolution, stellar collisions, and evolution of close black-hole binaries. Moreover, it can be easily modified so that the part of the code that uses GRAPE-6, could be replaced by another module that uses other accelerating-hardware such as the graphics processing units (GPUs).} 
{Appropriate choice of the free parameters give a good accuracy and speed for simulations of star clusters up to and beyond core collapse. Simulations of Plummer models consisting of equal-mass stars reached core collapse at $t \simeq 17$ half-mass relaxation times, which compares very well with existing results, while the cumulative relative error in the energy remained below $10^{-3}$. Comparisons with published results of other codes for the time of core collapse for different initial conditions, show excellent agreement. Simulations of King models with an initial mass-function, similar to those found in the literature, reached core collapse at $t \simeq 0.17 \pm 0.05$ half-mass relaxation times, which is slightly earlier than expected from previous work. Finally, the code accuracy becomes comparable to and even better than the accuracy of existing codes, when a number of close binary systems is dynamically created in a simulation. This is because of the high accuracy of the method that is used for close binary and multiple subsystems.}
{The code can be used for evolving star clusters containing equal-mass stars or star clusters with an initial mass function (IMF) containing an intermediate mass black hole (IMBH) at the center and/or a fraction of primordial binaries, which are systems of particular astrophysical interest.}


\maketitle

\section{Introduction}

Systems of numerous individual members interacting with each other are found on both small and large scales. Gas molecules interact with each other by means of electrostatic van der Waals forces, the same forces are responsible for the interactions and collisions of amino-acids that lead to the construction of protein molecules. On the other hand, planets and planetesimals interact with both each other and the central star in a planetary system, with the gravitational force. The gravitational force is also the interaction force between members of greater systems such as open or globular star-clusters, galaxies, and groups of galaxies.

N-body codes provide one way of simulating systems like those mentioned above. Those codes require calculations of all the forces between all the individual particles at every integration time step. This means that at every time-step, a total of $N(N-1) \approx N^2$ force calculations should be made, where $N$ is the number of interacting particles. From that, it is easy to conclude that N-body codes have time complexity $O(N^2)$ and require fast computers or even special purpose hardware, such as GRAPE systems, to simulate realistically large systems. This is why the evolution of N-body codes that could integrate these systems in a reasonable time, is closely connected to the hardware evolution.

The first attempt to perform a simulation, albeit without the use of a computer, but instead a system of 37 light bulbs represending two interacting model galaxies, was made by Holmberg \citet{Holmberg}. The first computer N-body simulations were developed by \citet{vonHoerner1,vonHoerner2} using initially 16 and later 25 particles. Later, as simulation methods and computers became faster, the number of particles increased to 100 \citep{Aarseth1963} and 250 \citep{Aarseth1968}. Even in these early attempts some of the physical processes that occur in true star clusters, such as mass segregation and formation of hard binaries, were found in the simulations, but simulating systems closer to realistic star clusters required improvements in both software, by developing efficient algorithms, and hardware.

A significant step in the evolution of N-body algorithms was the introduction of individual time steps: all particles do not share the same time step, but each one of them has its own, depending on its motion. In block-time-step schemes \citep{McMillan1985}, the time steps are quantized, usually as powers of 2, permitting blocks of particles to be updated simultaneously. This idea made the algorithms faster, because only the necessary calculations need to be made for a given accuracy. Today all sophisticated N-body codes such as \texttt{Starlab}\footnote{\url{http://www.ids.ias.edu/~starlab/} also see \url{http://muse.li} for a set of N-body simulation tools} \citep{Starlab1, Starlab2}, $\phi$GRAPE \citep{Harfst2008}, and Aarseth's \texttt{nbody}\footnote{\url{http://www.ast.cam.ac.uk/~sverre/web/pages/nbody.htm}} series (\texttt{nbody4}, \texttt{nbody6}, \texttt{nbody6++} \citep{FromNbody1to6,Nbody6++1,Nbody6++2}) use block time steps. Another milestone in the history of N-body simulations was the introduction of special purpose hardware, responsible only for the calculation of the gravitational forces. The GRAPE series of computers \citep{Grape4,Grape6} were designed to perform only this type of calculation. The next step was GRAPE-DR\footnote{\url{http://www.kfcr.jp/index-e.html}} \citep{GrapeDR}, which is a more general purpose hardware specialized for N-body codes which operates faster than a normal CPU. Simulations of realistically large star systems can be made using a single GRAPE machine attached to a fast host-computer. Today, GPUs are slowly replacing CPUs and the older GRAPEs in N-body simulations \citep{GPU1,GPU2,Chamomile}, because they are faster and considerably cheaper, than either CPUs or GRAPEs. 

In this paper, we assess the limitations of N-body codes with today's software and hardware power. First of all, N-body simulations are eithercollisionless or collisional. In collisionless simulations, the gravitational law is modified in close encounters to avoid collisions of particles and disencourage the dynamical formation of hard binary systems. The modified gravitational force acting on a single particle $i$ due to all other particles is calculated to be:

\begin{equation} \label{Newton}
 \vec{F}_{i} = \sum_{j\neq i} \frac{m_i m_j}{{(r_{ij}^2 + \epsilon^2)}^{3/2}}\vec{r_{ij}},
\end{equation}
\noindent where $\epsilon$ is the so-called softening parameter. The presence of close binary systems slows down the simulations, thus when handling large numbers of particles, one prefers to use collisionless simulations.

In simulations of collisional systems, the minimum number of stars that can be modeled is much lower and depends strongly on the fraction of stars that belong to binary systems in the initial configuration (primordial binaries). This is because the evolution of binary systems requires special treatment, including small time steps. \citet{Baumgardt2005} presented the results of simulations of unequal-mass star clusters containing up to $N=131072$ stars and an intermediate-mass black hole (IMBH) at the center. Those simulations were oparated for 12 Gyrs and no primordial binaries were included. In \citet{withPrimordial}, the results of simulations of star clusters with a total of $N=144179$ stars including $10\%$ primordial binaries were presented. The simulations were run for 100 Myrs.  From the above numbers, it is easy to conclude that although it is not possible to simulate a typical galaxy with all its stars $(N\approx 10^8-10^9)$, a simulation of a typical globular cluster $(N\approx 10^5-10^6)$ for timescales of the order of some Gyrs is possible, with today's computing power. If primordial binaries are not included, the evolution of a star cluster with up to $10^5$ stars up to and beyond core collapse is possible, using a GRAPE-6 machine attached to a fast host-computer. In the near future, using new algorithms and the coming generations of hardware, it should become possible to simulate the dense and massive centers of galaxies $(N\approx 10^6-10^7)$.

During the past few years, N-body simulations have provided a realistic tool for explaining the dynamics and structure of star clusters. \citet{Zwart1999} and \citet{Zwart2003} explained one way of forming IMBH in some globular clusters, by means of the process of runaway collisions of stars. \citet{Baumgardt2005} used detailed N-body simulations to discuss the structure of a globular cluster containing an IMBH, while \citet{Ejection} discussed the possibility of the ejection of supermassive black holes from galactic centers. \citet{PostNewtonian1} and \citet{GWaves} used post-Newtonian effects on their N-body simulations to study the dynamics and the gravitational radiation from IMBH-binaries in star clusters. Finally, \citet{segragation1} discussed the effects of primordial mass segregation on the evolution of globular clusters.

Here we introduce \texttt{Myriad}\footnote{Myriad is an ancient Greek unit for 10\,000, which is the order of magnitude of the size of a star cluster that can be evolved using the code, in reasonable time. In modern English, the word refers to an unspecified large number \url{http://en.wikipedia.org/wiki/Myriad}}, a new C++ N-body code for collisional and collisionless simulations. The code is loosely based on the instructive on-line books found in \textit{The Art of Computational Science} website\footnote{\url{http://www.artcompsci.org}}. It has many of the features of existing codes, such as the use of block time steps, the Hermite $4^{th}$ order integrator, GRAPE-6 support, and special treatment for binary and multiple sub-systems, but introduces a new structure and some new ideas especially in using GRAPE-6 to find neighboring stars and perturbers of binary systems. 

These ideas and a description of the basic structure of the code are presented in Sect. 2. In Sect. 2.2, the detailed steps of the code are described. More attention is given in describing the method for dynamical evolution of binary or multiple sub-systems that may appear in a simulation, in Sect. 2.2.3. In Sect. 3 we compare our preliminary results with the results of other codes. In the final two sections, some of the future directions of this code are presented with a discussion about the source of numerical errors and the CPU time needed for the simulations. The code uses the so-called N-body units, which, together with some basic equations for calculating the physical parameters of a star cluster are given in Appendix A.

\section{The new code}

\subsection{Code structure}

\begin{figure*}[th]
\centering
\includegraphics[width=10cm,height=8cm]{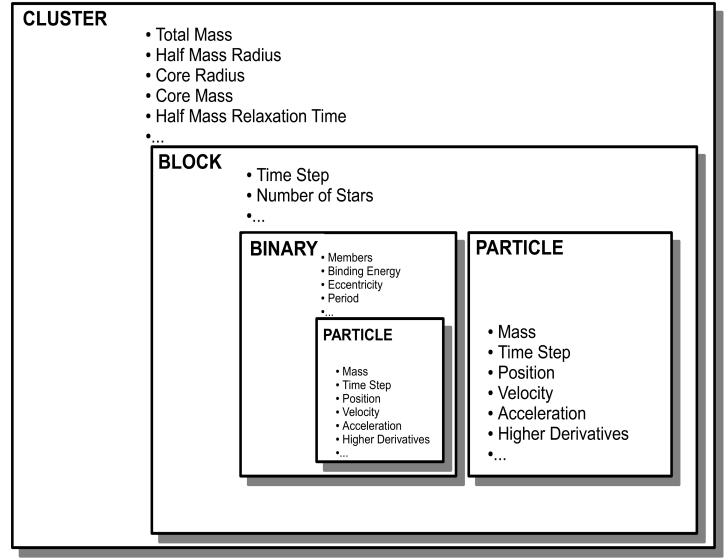}
\caption{Class hierarchy in \texttt{Myriad}. A class \textsc{cluster} is a set of several \textsc{block} classes. A \textsc{block} contains a variable number of \textsc{particle} and \textsc{binary} classes. A \textsc{binary} is a set of two or more \textsc{particle} classes. A class \textsc{particle} contains information (mass, position, velocity etc) for a single star and functions that act on them. The class \textsc{cluster} contains information about the system as a whole (total mass, energy, half-mass radius, core radius etc) and the required functions for calculating them.}
\label{fig:fig0}
\end{figure*}

\begin{figure*}[th]
\centering
\includegraphics[width=14cm,width=12cm]{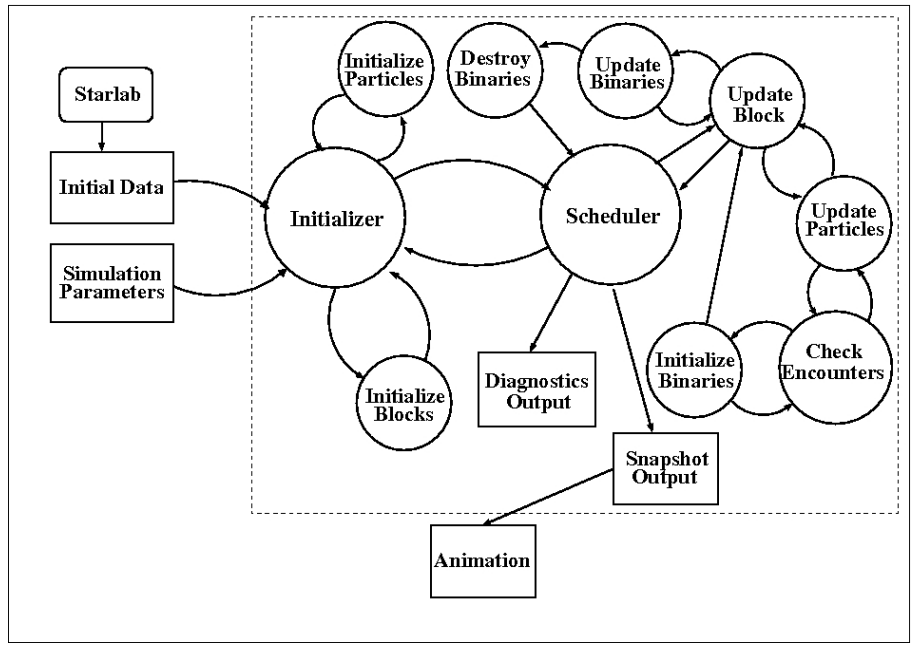}
\caption{Simplified graphical representation of \texttt{Myriad}. The arrows show the flow of the data in the code. Boxes represent input or output datafiles or outside applications, while circles represent sets of code-functions. \texttt{Myriad} applications are those inside the large dashed box. The other boxes and circles refer to satellite programs that create initial data or make animations from \texttt{Myriad}-output.}
\label{fig:fig1}
\end{figure*}

The core of \texttt{Myriad}\footnote{\texttt{Myriad} will be soon available for download on \url{http://www.astro.auth.gr/~simos/mediawiki-1.6.7/index.php/Myriad}} uses C++ classes to store and manipulate data. The simplest or lower-level class is the class \textsc{particle}, which contains all necessary information about a single star\footnote{A star is also referred as particle} (mass, radius, position, velocity, acceleration etc). The class \textsc{particle} also contains a set of functions that act on a star following its evolution in time, finding its neighbors, its time step and all the required data for the star itself. 
A set of \textsc{particle} classes that share the same time step at a given time, forms a higher-level class, the class \textsc{block}. As with the class \textsc{particle}, for the class \textsc{block} there is a set of functions that act on both the class itself and its \textsc{particle} members.

Finally, the complete set of \textsc{block} classes (it might only be one if all stars evolve with the same time step) forms a \textsc{cluster}, which is the highest-level class. The \textsc{cluster}-functions, act on all \textsc{particle} classes of the system, and mainly finding information about the cluster of stars itself (total mass, center of mass, escapers, binaries etc).
There is also another class that is a member of a class \textsc{block}, the class \textsc{binary}, which consists of two or more \textsc{particle} classes representing stars that lie close to each other in space. The \textsc{binary} class has the required functions for the initialization of a binary or multiple star-system, its evolution in time, and its termination, when necessary, and also returns the required data back to the \textsc{cluster}.

Figure \ref{fig:fig0} shows the hierarchy of classes in the code described above. Figure \ref{fig:fig1} is a simplified graphical representation of \texttt{Myriad}. Initial data are provided to \texttt{Myriad} from a single data file. This initial data files can be constructed by the \texttt{Starlab} package. The main output of \texttt{Myriad} are snapshots of the evolving system at every output time step $dt_{\rm out}$ and information about the whole cluster (total mass, core radius, half-mass radius etc) at every diagnostic time step $dt_{\rm diag}$.

\begin{figure*}[!th]
\centering
\includegraphics[width=12cm,width=12cm]{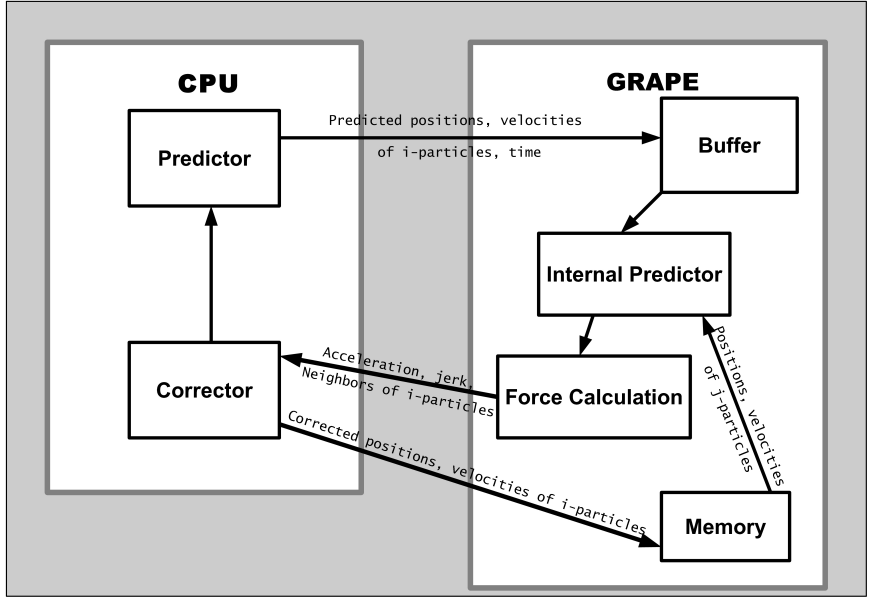}
\caption{Schematic data flow between the CPU and the GRAPE-6 for the Hermite $4^{th}$-order integrator. See text for description.}
\label{fig:fig2}
\end{figure*}

\subsection{Integration method}
The integration method used in \texttt{Myriad} is the Hermite $4^{th}$ order (H4) \citep{Hermite4} predictor-corrector scheme (PEC: Predict-Evaluate force-Correct) with block time steps. Accelerations and their derivatives (jerks) are computed using GRAPE-6. Binaries, close encounters, and multiple sub-systems that require small time steps are detected using GRAPE-6 and evolved using the time-symmetric Hermite $4^{th}$ order integrator \citep{tsHermite4}, where the prediction (P) step is the same as in the simple H4, but the force evaluation (E) and correction (C) steps are repeated 3 times $(P(EC)^3)$.

For each \textsc{block} $i$, ${t_{i}}_{\rm c}$ is the current time of its \textsc{particle} classes, $dt_{i}$ is the time step, and $t_{i \rm f} = t_{i \rm c} + dt_{i}$ is the time after a single update.
 
The integration procedure consists of the following steps (the time complexity of the slowest steps is displayed in parenthesis):
\begin{enumerate}

\item We define the universal current time $t_{\rm c}$ and the time after the step is completed $t_{\rm f}$ to be equal to the time and time forward of \textsc{block} 0, the \textsc{block} containing stars with the minimum time step, respectively
\begin{equation}\label{time_now}
	t_{\rm c} = t_{0 \rm c},
\end{equation}
\begin{equation}\label{time_f}
        t_{\rm f} = t_{0 \rm f}.
\end{equation}

\item We determine which \textsc{blocks} have time forward equal to $t_{\rm f}$ and label them as requiring update.

\item We update each labeled \textsc{block} to its time forward. \\
To update a single \textsc{block} with $m$ members ($N$ being the total number of stars):
\begin{enumerate}
\item If \textsc{binary}\footnote{A \textsc{binary} can contain two or more stars.} classes exist, update them to $t_{\rm f}$ \\
To update a \textsc{binary} class with $k$ members:
\begin{enumerate}
\item We determine the time step $dt_{\rm b}$ for the $P(EC)^3$ integrator.
\item We predict the position and velocity of each member $\Big( O(k) \Big)$.

\item We calculate the forces between binary-members on the CPU $\Big( O(k^2) \Big)$.
\item We calculate the perturbation forces from $p$ neighboring stars on the CPU $\Big( O(kp) \Big)$.
\item We then correct the position and velocity of each member $\Big( O(k) \Big)$.
\suspend{enumerate}
Processes (iii) to (v) are repeated $n=3$ times.
\resume{enumerate}
\item We update the current time $t_{\rm c} = t_{\rm c} + dt_{\rm b}$ of the \textsc{binary}.
\item We continue from (i) until $t_{\rm c}$ equals $t_{\rm f}$.
\end{enumerate}

\item We predict the positions and velocities of all the particles of the \textsc{block}, including normal stars and ``virtual'' stars that represent the centers of mass of binary or multiple sub-systems $\Big( O(m) \Big)$.
\item We evaluate the accelerations and their derivatives for all the particles of the \textsc{block} using the GRAPE-6 hardware $\Big( O(m\log N) \Big)$.
\item We find the new time step for each particle of the \textsc{block}. This will be used for the next update $\Big( O(m) \Big)$.
\item We check for encounters between single stars or binaries.
\item We correct the positions and velocities of all the particles of the \textsc{block} $\Big( O(m) \Big)$.
\item If an encounter is detected in step (e), we construct the \textsc{binary}, and replace its members with a ``virtual'' star represending their center of mass.
\item We check the \textsc{binary} classes. If a \textsc{binary} is terminated, replace its center of mass in the N-body code, with its former members and erase it.
\item We then send the corrected values to GRAPE-6 memory.

\end{enumerate}

\item We move particles between \textsc{block} classes according to their new time step.

\item We update $t_{\rm c}$ and $t_{\rm f}$ for each updated \textsc{block}. 

\item We continue with step 1.

\end{enumerate}

From the above, it is obvious that the speed of the code depends on:
\begin{enumerate}
 \item The total number of stars $N$.
 \item The number of \textsc{block} classes and the average number of stars per \textsc{block}.
 \item The average number $p$ of stars responsible for significant perturbations on a binary system (perturbers).
\end{enumerate}

Figure \ref{fig:fig2} shows how data flows between the CPU and the GRAPE-6 and which parts of the calculation are performed on the CPU and the GRAPE-6. Initially positions and velocities, and estimates of both the accelerations and the jerks of all particles are stored in GRAPE-6 memory (This procedure is omitted in the graph). Then, a prediction is made for a set of i-particles using the CPU. The predicted values and the time of each particle is sent to the GRAPE-6 buffer, which directly sends them to the internal predictor of the GRAPE-6. The internal predictor takes the data for the rest j-particles from the GRAPE-6 memory and predicts their positions and velocities at the time of the i-particles. The GRAPE-6 then computes the acceleration, its derivative (jerk), and the nearest neighbors for the i-particles and returns them to the corrector operating on the CPU. The corrector calculates the corrected values of position and velocity for the i-particles and sends the new data to the GRAPE-6 memory. A new set of i-particles then, is sent to the predictor and the procedure continues until all particles are evolved to the expected time. Data flow between the CPU and the GRAPE-6 occurs three times per time step. The most time-consuming calculation, the calculation of the gravitational forces ($O(N\log(N))$), performed on the GRAPE-6. The rest of the calculations are of order $O(N)$. If the force-calculation were to be performed on the CPU, then it would be far more time-expensive as it would scale as $O(N^2)$.

\subsubsection{Block time steps}

\begin{figure*}[!t]
\centering
\includegraphics[width=10cm,height=8cm]{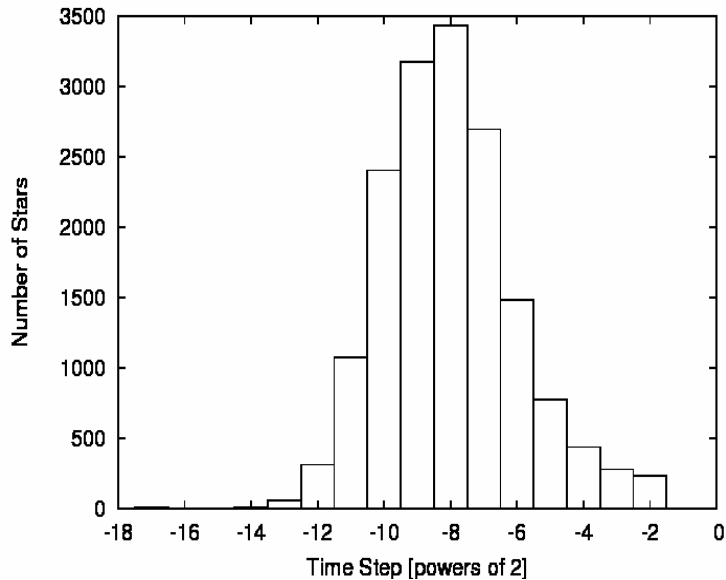}
\caption{Distribution of equal mass particles in different time steps (powers of 2). Aarseth's criterion is used for the time step with $\eta = 0.01$. A small number of particles lies in time step $2^{-14}$. This is the smallest allowed time step for the N-body code. Stars with this time step are candidates for close encounters or virtual particles representing the centers of mass of binary or multiple sub-systems} 
\label{fig:fig6}
\end{figure*}

The use of block time steps have proven to be an ideal method for N-body simulations. The advantage of block time steps over constant (shared) time steps for all particles is that a smaller number of operations is needed to achieve a given accuracy. The advantage over individual time steps is that particles are monitored with time in groups and not individually. The implementation of block steps is ideal with GRAPE-6, because it lessens the communication time between the host and GRAPE-6. In individual time-step algorithms, a single particle is sent to GRAPE-6 everytime, while in block time step algorithms, groups of particles, that share the same time step are sent simultaneously to GRAPE-6 for force evaluation.

At $t=0$, all particles, unless there are binaries in the initial configuration, share the same time step, which is the smallest time step allowed by the N-body code. This time step is given by \citep{AarsethBook}
\begin{equation}\label{dt_min}
 Dt_{\rm min} = 0.04 \Big( \frac{\eta_{\rm I}}{0.02} \Big)^{1/2} \Big( \frac{R^3_{\rm cl}}{\bar{m}} \Big)^{1/2},
\end{equation}
where $\bar{m}$ is the mean mass of the system, $\eta_I$ the initial accuracy parameter with typical value $0.01$, and $R_{\rm cl}$ is the close encounter distance
\begin{equation} \label{R_cl}
 R_{\rm cl} = \frac{2G\bar{m}}{\sigma^2},
\end{equation}
where $G$ is the gravitational constant, $\sigma$ is the rms velocity dispersion given by
\begin{equation} \label{sigma}
 \sigma^2 = \frac{G N \bar{m}}{2R_{\rm v}}
\end{equation}
at equilibrium, $N$ is the number of stars, and $R_{\rm V}$ is the virial radius of the cluster
\begin{equation} \label{virial}
 R_{\rm V} = G \frac{N^2 \bar{m}}{|2U|},
\end{equation}
where $U$ is the total potential energy.

The time step $Dt_{\rm min}$ is rounded to the closest power of 2 and the resulting $\Delta t_{\rm min}$ is the time step of \textsc{block} 0 and the particles that belong to it.

Each time a particle $i$ is updated, its next time step is calculated according to Aarseth's criterion \citep{AarsethBook}
\begin{equation}\label{Aarseth}
dt_{i,1} = \sqrt{ \eta  \frac{ |\bf{a}_{i,1}| |\bf{\ddot{a}}_{i,1}| + |\bf{\dot{a}}_{i,1}|^2} { |\bf{\dot{a}}_{i,1}| |\bf{\dddot{a}}_{i,1}|  + |\bf{\ddot{a}}_{i,1}|^2} },
\end{equation}
where $\eta$ is the accuracy parameter with a typical value $0.01$, and $\bf{a}_{i,1}$, $\bf{\dot{a}}_{i,1}$,  $\bf{\ddot{a}}_{i,1}$, and $\bf{\dddot{a}}_{i,1}$ are the acceleration of particle $i$ and its time derivatives, calculated in next section. 
Subscript ``$1$'' refers to the values of the parameters at the end of the time step (final values), while subscript ``$0$'' refers to the values at the beginning (initial values). 

The second and third time derivatives of the acceleration at the end of the time step are given by
\begin{equation} \label{a_ddot1}
\ddot{\textbf{a}}_{i,1} = \ddot{\textbf{a}}_{i,0} + \Delta t_{i,0} \dddot{\textbf{a}}_{i,0}
\end{equation}
and 
\begin{equation}
\dddot{\textbf{a}}_{i,1} = \dddot{\textbf{a}}_{i,0},
\end{equation}
where $\Delta t_{i,0}$ is the previous time step of particle $i$. 

Time steps are quantized according to the rule $\Delta t_{i,1} = 2^{n}$, where $n$ is a negative integer defined in such a way that
\begin{equation} \label{Delta_t}
2^n \leq dt_{i,1} < 2^{n+1}.
\end{equation}
In this way, all particles have time steps that are powers of 2. All particles with the same time step are grouped in a \textsc{block} so that they are updated simultaneously. The time step of ``virtual'' particles replacing binaries or multiple systems (see below) is set to be $\Delta t_{\rm min}$. In addition, $\Delta t_{\rm  min}$ is the time step of particles that are relatively close to and approaching each other rapidly. To improve the efficiency of the code, every new time step cannot be longer than 2 times its previous value, jumps to higher time steps with step greater than 2 are not allowed.

The diagram of Figure \ref{fig:fig6} shows an example of a particle distribution in different time steps in a cluster of 16\,384 ($2^{14}$) stars.

\subsubsection{The Hermite $4^{th}$ order predictor-corrector scheme}

\begin{figure*}[!ht]
\centering
\includegraphics[width=18cm,height=14cm]{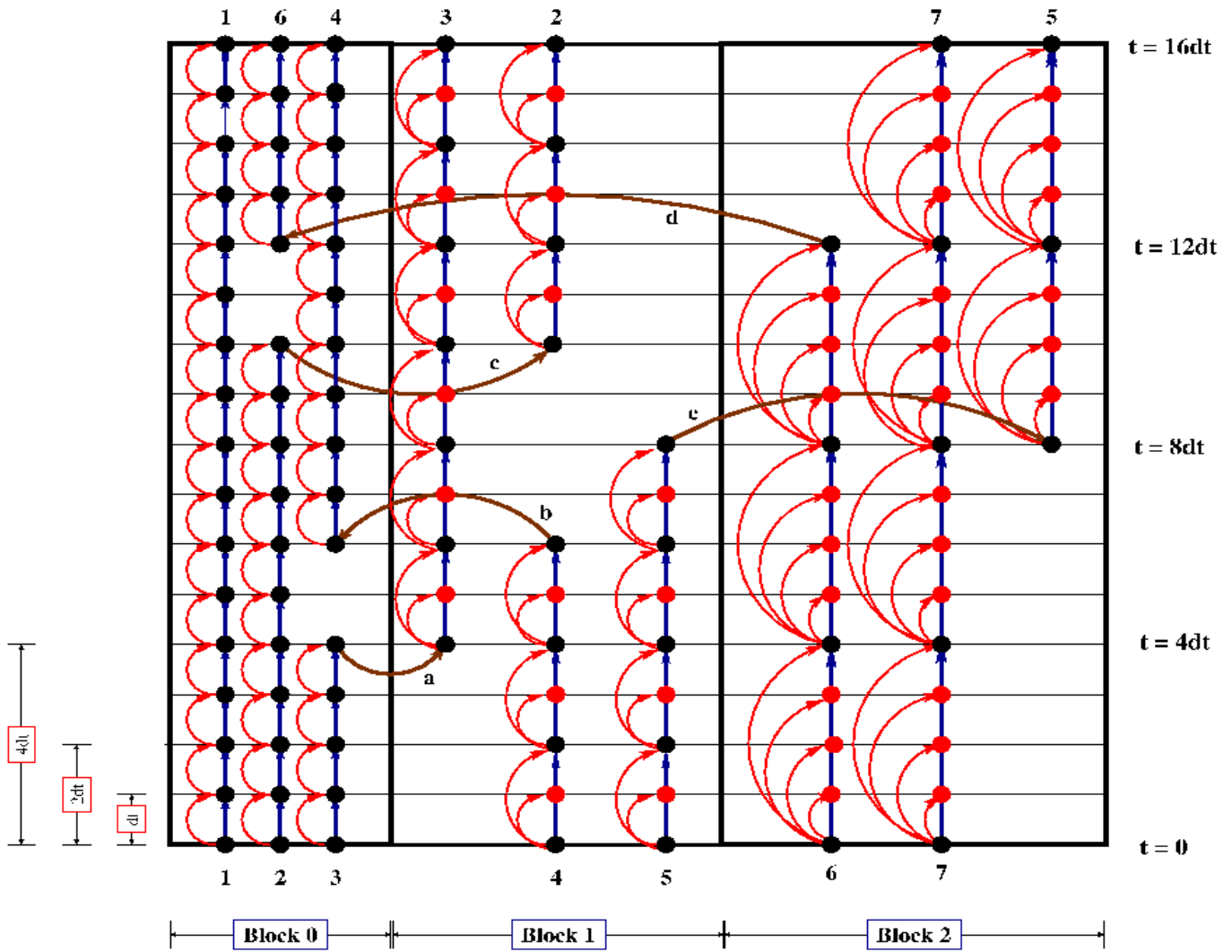}
\caption{Representation of Hermite $4^{th}$ order with block time steps. In the figure, a set of 7 particles distributed in 3 time-blocks is shown. See text for description.}
\label{fig:fig4}
\end{figure*}

The H4 scheme is the integrator used in many N-body codes. The most important feature of this integrator is that it needs only one calculation of the acceleration and its derivatives (force calculation) in every time step. This makes H4 superior to other popular integrators of the same accuracy, which require more than one force calculations per time step. A Runge-Kutta $4^{th}$ order algorithm evaluates the force three times every time step, which makes this integrator much slower than H4 and consequently inappropriate for N-body simulations with large number of particles.

Another reason for choosing the H4 integrator is that GRAPE-6 is specifically designed for use with this kind of integrator. This is because GRAPE-6 returns the acceleration and its first derivative for each particle, which are the only data H4 needs. One may use a higher order integrator of the Hermite family \citep{Hermite6}, but this would require extra calculations to be made of higher order derivatives of the acceleration on the CPU. Higher accuracy, also is not necessary for this kind of problem.

The integration steps for advancing a particle forward in time with the H4 scheme are given in the following lines (see also \citet{Hermite4}).

We assume that $t_{i,0}$ is the current time of the particle, $t_{i,1} = t_{i,0} + \Delta t_{i,0}$ is the time after a step is taken, and $\Delta t_{i,0}$ is the current time step.
\begin{enumerate}
 \item We predict the position and velocity of the particle using already known data:
\begin{align}
&\textbf{x}_{i,\rm pred} = \textbf{x}_{i,0} + \Delta t_{i,0} \textbf{v}_{i,0} + \frac{1}{2} {\Delta t}^2_{i,0} \textbf{a}_{i,0} + \frac{1}{6} {\Delta t}^3 \dot{\textbf{a}}_{i,0}  \label{x_p} \\
&\textbf{v}_{i,\rm pred} = \textbf{v}_{i,0} + \Delta t_{i,0} \textbf{a}_{i,0} + \frac{1}{2} {\Delta t}^2_{i,0} \dot{\textbf{a}}_{i,0} \label{v_p}
\end{align}
This operation has a time complexity of $O(1)$.
\item We evaluate the acceleration $\textbf{a}_i$ and its derivative (jerk) $\dot{\textbf{a}}_i$ of the particle. To achieve this, we have first to predict the positions and velocities of all the other particles to time $t_{\rm f}$. The acceleration and jerk are evaluated on either the GRAPE-6 or the CPU using
\begin{eqnarray} 
&&\textbf{a}_{i,1} = \sum_{j\neq i} G m_j \frac{\textbf{r}_{ij}}{ r_{ij}^3 } \label{a1}, \\
&&\dot{\textbf{a}}_{i,1} = \sum_{j \neq i} G m_j \left[  \frac{ \textbf{v}_{ij} }{r_{ij}^3} - \frac{3 (\textbf{v}_{ij} \cdotp \textbf{r}_{ij}) \textbf{r}_{ij}} {r_{ij}^5}  \right] \label{a_dot},
\end{eqnarray} 
where 
\begin{equation}
 \textbf{r}_{ij} = \textbf{x}_{j,\rm pred} - \textbf{x}_{i,\rm pred},
\end{equation}
\begin{equation}
 \textbf{v}_{ij} = \textbf{v}_{j,\rm pred} - \textbf{v}_{i,\rm pred}.
\end{equation}

If softening $\epsilon$ is included in the calculations (for collisionless simulations), then the denominators of Eqs. (\ref{a1}) and (\ref{a_dot}) are replaced with $({r}_{ij}^2 + \epsilon^2)^{(3/2)}$ and $({r}_{ij}^2 + \epsilon^2)^{(5/2)}$, respectively.

We also, evaluate numerically higher order derivatives of the acceleration using
\begin{align} 
&\ddot{\textbf{a}}_{i,0} = \frac{ -6 (\textbf{a}_{i,0} - \textbf{a}_{i,1} ) - \Delta t_i (4 \dot{\textbf{a}}_{i,0} + 2 \dot{\textbf{a}}_{i,1})}{ \Delta t_i^2 } \label{a_ddot}, \\
&\dddot{\textbf{a}}_{i,0} = \frac{ 12 ( \textbf{a}_{i,0} - \textbf{a}_{i,1}  ) + 6 \Delta t_i (\dot{\textbf{a}}_{i,0} + \dot{\textbf{a}}_{i,1} ) }{ \Delta t_i^3} \label{a_dddot}.
\end{align}
The evaluation of $\textbf{a}_{i,1}$ and $\dot{\textbf{a}}_{i,1}$ has time complexity $O(N)$, when it is performed on the CPU and $O(\log N)$, when it is performed on the GRAPE-6. The other calculations have complexity $O(1)$.

\item We correct the position and velocity of the particle using higher order terms
\begin{equation} \label{x_c}
 \textbf{x}_{i,\rm cor} = \textbf{x}_{i,\rm pred} + \frac{\Delta t_i^4}{24} \ddot{\textbf{a}}_{i,0} + \frac{\Delta t_i^5}{120} \dddot{\textbf{a}}_{i,0}, 
\end{equation}
\begin{equation} \label{v_c}
 \textbf{v}_{i, \rm cor} = \textbf{v}_{i,\rm pred} + \frac{\Delta t_i^3}{6} \ddot{\textbf{a}}_{i,0} + \frac{\Delta t_i^4}{24} \dddot{\textbf{a}}_{i,0}.
\end{equation}
The complexity of this operation is $O(1)$.
\end{enumerate}

Figure \ref{fig:fig4} shows how particles distributed in 3 time-blocks are updated using the H4 scheme. We note that in a typical star cluster, the number of time-blocks usually used is 10-15. All particles lie initially at the bottom, where $t=0$. Initially block 0 contains 3 particles, while blocks 1 and 2 each contain 2 particles. A black dot represents a particle after the correction step (a straight blue arrow represents the correction step), while a red dot represents a particle after a prediction step (a curved red arrow represents the prediction step). Force evaluation is necessary for every particle, only before a correction step, i.e., force evaluation happens for a particle only when it is black. For the force evaluation of a particle, a prediction of all other particles to the current time of the particle is needed. It is obvious that particles in block 0 are considered for predict-evaluate-and-correct every $dt$. Particles of block 1 predict every $dt$, but subjet to evaluate-and-correct every $2dt$. Finally, particles of block 2 predict every $dt$ and evaluate-and-correct every $4dt$. The curved brown arrows show how particles are allowed to move between blocks depending on the time step calculated using Eq. ($\ref{Aarseth}$). A jump to a higher block is allowed only if it doubles the time step of a particle (particle 3 jumps from block 0 to block 1). In contrast, a jump to any lower block is allowed if necessary (particle 6 jumps immediately from block 2 to block 0).

\subsubsection{Binaries and multiples} 

During the evolution of the system, close encounters between stars may occur. Close binaries also may be formed dynamically, after three-body encounters. Those subsystems require a small time step to be integrated accurately in time and this time step may be much shorter than the shorterst time step $\Delta t_{\rm min}$ used in the N-body code. This means that this subsystem has to be specially treated to ensure that it evolves in time with the required accuracy and speed. The special treatment used for close encounters, binaries, and small multiple systems is described in the following lines.

The time-symmetric Hermite $4^{th}$ order integrator \citep{tsHermite4} is the numerical integrator for binaries, multiples, and close encounters. This integrator is more accurate than the simple H4 integrator with the penalty of being  a few times slower. Time-symmetry is achieved by applying the force evaluation-correction (EC) steps $n$ times, after  the prediction $(P(EC)^n)$ and by appropriate choice of the time step. The value $n=3$ is used in the code because this is sufficient to achieve time-symmetry. Time-symmetry is responsible for transforming the H4 scheme into a symplectic integrator, which is more appropriate for the time evolution of small systems such as binaries. Symplectic integrators \citep{symplectic1} conserve the characteristics of the integrated system (energy, semimajor axis, eccentricity, period) with high accuracy, paying of course the penalty of time. In reality, when using symplectic integrators the energy of the integrated system is bounded and changes periodically with time. In contrast, non-symplectic integrators are faster, but the energy of the integrated system increases monotonically with time. A brief description of the integration method for binaries and multiple sub-systems is given at the beginning of Sect. 2.2, between items (i) and (vii). 

The choice of the time step for the $(P(EC)^3)$ integrator is important to preserve time-symmetry. The time step at each time $t$ must be such that if the time were to be reversed and the particles move backwards, the time step choice would remain the same. We followed the logic of the algorithm presented in \citet{time_symmetric1} for the choice of the time step. According to this algorithm, the time step is only allowed to either increase or decrease by a factor of 2 or remain the same with respect to its last value and an increase in the time step is allowed only at even times. In this way, the time step remains symmetric and quantized as a power of 2. The current time is characterized as either even or odd with respect to the last time step taken by the system. We apply this algorithm using as a time step criterion the shortest among the time steps $dt_{{\rm b}_i}$ of the members of the \textsc{binary}

\begin{equation} \label{dt_b}
 dt_{{\rm b}_i}  = \eta_{\rm b} \frac{ |\bf{a}_{i}| } { |\bf{\dot{a}}_{i}| },
\end{equation}
where $\eta_{\rm b}$ is a parameter to control the accuracy, and $\bf{a}_{i}$ and $\bf{\dot{a}}_{i}$ are the acceleration and its derivative for particle $i$, respectively.

At each time $t$, we begin by finding whether $t$ is even or odd with respect to the last time step, $\Delta t_{\rm old}$, taken. If $t$ is odd, we maintain the same time step and by using $\Delta t_{\rm old}$ we determine the time step at both the beginning $dt_{\rm 0}$ and the end $dt_{\rm 1}$ of that time step and find their average
\begin{equation} \label{criterion}
 dt = \frac{dt_{\rm 0} + dt_{\rm 1}}{2}.
\end{equation}
We constrain the time step used $\Delta t_{\rm old}$ to be shorter than $dt$ and longer than $dt/2$. If it satisfies this condition, then the chosen time step is the one to use in the integration. If not, then the time step must be halved. In the case of even time, we follow the same procedure, testing whether the doubling of the previous time step satisfies the criterion of being smaller than $dt$ and greater than $dt/2$, where $dt$ is given by Eq. (\ref{criterion}). If the criterion is satisfied, then $2\Delta t_{\rm old}$ is the time step for the next step. If not, then we follow the same logic using $\Delta t_{\rm old}$ as a test time step and follow exactly the same steps as if $t$ was odd.

For this algorithm to work properly, an appropriate choice for $\eta_{\rm b}$ must be made. This choice needs to be small enough for $dt$ not to change by more than a factor of 2 with respect to its previous value. In addition, $\eta_{\rm b}$ is responsible for providing a smooth transition to the time step from the N-body code to the binary module of the code, to ensure that a particle that has either just joined a binary system or just left it does not change its time step by more than a factor of 2. This is required because major changes in the time step introduce significant errors in the integration. To guarantee this behavior, $\eta_{\rm b}$ is calculated for every \textsc{binary} at the time of its formation. To do this, we determine the particle $i$ with the smallest value of $|\bf{a}_{i}|/ |\bf{\dot{a}}_{i}|$ and using Eq. (\ref{dt_b}) we determine $\eta_{\rm b}$ by requiring that $dt = Dt_{\rm min}/2$, which is given by Eq. (\ref{dt_min}) and is the smallest time step of the N-body code; it is also the last time step taken by all the future members of the \textsc{binary} or the next time step for all the particles that leave a \textsc{binary}. In the case of a \textsc{binary} with more than 2 members, we use a shared time step for all its members. To avoid major changes in the time step criterion, we use an even smaller value for $\eta_{\rm b}$. The problem here is that if a single particle escapes for example from a triple system, leaving behind a close binary system, the time step of the escaping particle is forced to jump to $Dt_{\rm min}$ even if it was much smaller when the particle was still a member of the triple. This necessary change in the time step may introduce errors in the integration. To solve this problem, an individual time step must be used for multiple subsystems. In this way, all members of the \textsc{binary} would have their own time steps and an escaping particle would increase its time step by just a factor of 2.

Finding the binary systems in an N-body system is a slow procedure with time complexity $O(N^2)$, unless all \textsc{particle} classes are members of a linked list, as in a binary search tree, or GRAPE-6 is used for acceleration. Here we use the neighbor-module of GRAPE-6 to search more rapidly for binaries and close encounters. Each star $i$ with mass $m_i$ is associated with a distance 
\begin{equation}
R_{{\rm h}_i} = 5 \sqrt{\frac{N}{2} (m_i + m_{\rm g})} R_{\rm cl},
\end{equation}
where $m_{\rm g}$ is the mass of the most massive star of the system, $N$ the total number of stars, and $R_{\rm cl}$ is given by Eq. (\ref{R_cl}). During the calculation of the force for the star $i$, this distance is sent to the neighbor-module of GRAPE-6, which returns the identities of the stars that lie in distances $r\leq R_{{\rm h}_i}$ from the star $i$. Those stars are considered as ``neighbors'' of the star $i$. Then, a search between the neighbors is performed to find out whether one or more of them is about to have a close encounter or to form a close binary system with star $i$. The criterion for a close encounter between the star $i$ and its neighbor $j$ is
\begin{equation} \label{R_crit}
R_{ij} \leq R_{{\rm crit}_{ij}} = \sqrt{\frac{N}{2} (m_i + m_j)} R_{\rm cl},
\end{equation}
\begin{equation}
 V_{ij} < 0,
\end{equation}
where $R_{ij}$ is the separation between stars $i$ and $j$, $m_i$, and $m_j$ their masses, $N$ is the total number of stars, and $V_{ij}$ is the relative velocity of the two stars. The choice of the distance $R_{{\rm crit}_{ij}}$ comes from our finding that in an equal-mass system, if the distance between two stars is more than $R_{\rm cl}$, then they can be integrated accurately using the H4 integrator with a time step greater than or equal to $Dt_{\rm min}$, given by Eq. (\ref{dt_min}), which is the shortest time step allowed by the block time-step scheme. If the separation is smaller than $R_{\rm cl}$, then a smaller time step is needed, so integrating the two particles in time with acceptable accuracy, needs some special treatment. The factor $\sqrt{\frac{N}{2} (m_i + m_j)}$ in Eq. (\ref{R_crit}) is used to generalize the criterion for a non-equal-mass system. Finally, the distance $R_{{\rm h}_i}$ that is sent to GRAPE-6 during the calculation of the force for star $i$, is simply 5 times the critical distance between this star and the most massive star of the system, to ensure that every close encounter between two stars is detected.
 
If the above criteria are fulfilled, a \textsc{binary} with the two \textsc{particle} classes as members is created. The two \textsc{particle} classes are removed from the N-body system and replaced by a new \textsc{particle} that corresponds to a ``virtual'' star represending the center of mass of the removed stars. This \textquotedblleft virtual\textquotedblright \, star has mass
\begin{equation} \label{Tmass}
 M = m_i + m_j
\end{equation}
and is placed at position 
\begin{equation}
 \textbf{r}_{\rm cm} = \frac{m_i\textbf{r}_i + m_j\textbf{r}_j}{m_i+m_j}
\end{equation}
with velocity
\begin{equation}
 \textbf{v}_{\rm cm} = \frac{m_i\textbf{v}_i + m_j\textbf{v}_j}{m_i+m_j}
\end{equation}
and behaves as if it were a normal star in the rest-frame of the N-body system.

A third star $k$ at distance $R_k$ from the center of mass of the binary, becomes a member of the binary transforming it to a triple system, if one of the criteria below is satisfied:
\begin{enumerate}
\item In all cases, if the distance of the star $k$ from at least one of the members of the binary is smaller than $2/3 R_{\rm cl}$
\item If the relative motion of the two stars is a simple close encounter ($e>1$, where $e$ the eccentricity of the system), when
\begin{equation}
 R_k \leq \sqrt{\frac{N}{2} (M + m_k)} R_{\rm cl},
\end{equation}
where $M$ is the total mass of the binary and $m_k$ the mass of the third star.
\item If the relative motion of the two stars is a close binary ($e<1$), then the dimensionless perturbation $\gamma_{k}$, acting on the binary from star $k$
\begin{equation}\label{gamma}
 \gamma_k = M_2 \Big( \frac{R_{\rm b}}{R_k} \Big)^3
\end{equation}
becomes greater than a critical value, i.e.,
\begin{equation}
\gamma_k \geq \gamma_{\rm crit}.
\end{equation}
\end{enumerate}
In Eq. (\ref{gamma}), $M_2 = 2\bar{m}/M$, where $M$ is given by Eq. ($\ref{Tmass}$), $\bar{m}$ is the mean mass of the system, $R_{\rm b}$ is the size of the binary defined to be either its semimajor axis for hard binaries (i.e., binaries of binding energy greater than the average energy of a particle) or the distance $R$ between its members in the case of soft binaries (i.e., binaries with binding energy lower than the average energy of a particle). 

The same criteria are used when a fourth particle comes close enough to a triple system. The critical value of $\gamma$ for soft binaries is $\gamma_{\rm crit} = 0.125$. For hard binaries, we use the value $\gamma_{\rm crit} = 0.015625$. In an equal-mass system, according to Eq. (\ref{gamma}), the first value corresponds to a distance $R_k = 2 R_{\rm b}$ for the third star, while the second to a distance $R_k = 4 R_{\rm b}$. This means that in an equal-mass system, a third star would become a member of a soft binary when it lies at a distance twice the size of the binary. If the binary is hard, the critical distance is 4 times the size of the binary. 

Another way of forming multiple sub-systems with number of stars $n\geq4$ is when two binaries or multiples approach each other. When the distance between their centers of mass becomes smaller than three times the sum of their sizes, then the two binaries merge forming a larger system where all particles interact strongly with each other. The size of a binary is defined to be its semimajor axis, for hard binaries or the distance between its members, for soft binaries. 

Termination of a triple, quadruple, or higher order sub-system occurs when the following three criteria are met:
\begin{enumerate}
\item When the dimensionless perturbation $\gamma_j$ acting on the particle-member of the sub-system $j$ on the inner binary, becomes less than $\gamma_{\rm crit}$
\begin{equation}
 \gamma_j < \gamma_{\rm crit}.
\end{equation}
\item The relative velocity of particle $j$, $v_j$, with respect to the inner binary is positive, i.e., particle $j$ moves outwards.  
\item The distance of particle $j$ from any of the other particles/members of the multiple system, is greater than $R_{\rm cl}$.
\end{enumerate}
The so-called ``inner binary'' is the hardest double sub-system inside the multiple (the one with the greatest binding energy).

In some rare cases, two particles-members of a multiple system with number of stars $n\geq4$ are removed together from the system and form a close binary. 

Termination of a simple binary occurs when the distance between its members becomes greater than $R_{\rm cl}$. Hard binaries are not terminated, unless another particle becomes a member and interacts strongly with their members. The outcome of these interactions may be a softer or harder binary, a collision, or even a dissolved system.

\begin{figure*}[ht]
\centering
\includegraphics[width=6.5cm,height=8cm]{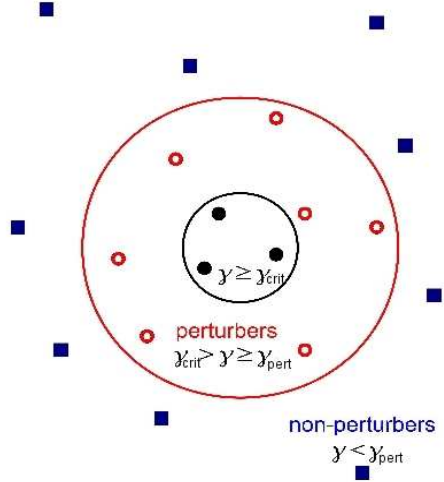}
\caption{\textbf{Perturbers}. Stars with $\gamma \geq \gamma_{\rm pert}$ are considered as perturbers of a binary system. Those are the stars represented as empty (red) circles in the above image. The filled (black) circles represent the members of the binary or multiple sub-system. If the dimensionless perturbation $\gamma$, given by Eq. (\ref{gamma}), becomes greater than $\gamma_{\rm crit}$, then the star is included into the binary or multiple sub-system. More stars could be added forming a multiple system with up to $\sim10$ stars. Those systems are usually dispersed quickly. On the other hand, a star with  $\gamma$ becoming smaller than $\gamma_{\rm pert}$ is no longer a perturber. Those are the stars represented with filled (blue) squares. For an equal mass system, we typically set $\gamma_{\rm pert} \sim 10^{-7}$ and $\gamma_{\rm crit} = 0.125$ if the inner binary is considered as ``soft'', while $\gamma_{\rm crit} = 0.015625$ is assumed if it is considered as ``hard''.}
\label{fig:fig5}
\end{figure*}

\paragraph{Collisions:}

Another rare case for termination of a binary is a collision between its members. A collision between two stars occurs when their distance $r_{ij}$
\begin{equation}
 r_{ij} \leq R_i + R_j,
\end{equation}
where $R_i$, $R_j$ the radii of the two stars.

The stellar radius $R_i$ for a main sequence star is given by
\begin{equation}
 R_i = R_\odot \Big( \frac{m_i}{M_\odot} \Big)^{a},
\end{equation}
where $m_i$ is the mass of the star, $R_\odot$ is the radius, and $M_\odot$ is the mass of the sun. The parameter $a$ depends on the mass $m_i$ \citep{BowersAndDeeming} such that
\begin{equation}
  a = \left\{
              \begin{array}{ll}
                   1 & (m_i \leq 0.5 M_\odot)\\
                   0.73 & ( m_i > 0.5 M_\odot)
              \end{array}
       \right.  
\end{equation}

If the star is a black hole, then $R_i$ corresponds to its Schwarchild radius
\begin{equation}
 R_i = \frac{2m_i}{c^2},
\end{equation}
where $c$ is the speed of light.

\paragraph{Perturbers:} 
The internal motion of a binary or multiple sub-system is controlled by the internal forces acting between its member stars. External perturbations from stars passing close to the center of mass of the sub-system are also included. A star $k$ is considered as a perturber of a binary system if the dimensionless perturbation $\gamma_k$, acting from the star to the binary, is greater than a critical value
\begin{equation} \label{gamma_pert1}
\gamma_k \geq \gamma_{\rm pert}.
\end{equation}
A typical value of $10^{-7}$ is used for $\gamma_{\rm pert}$, but this value is usually modified to limit the total number of perturbers for a binary system according to the total number of stars. During the force calculation, the external perturbations for every star-member of a binary or multiple system are calculated and added to their internal values.
The perturbation in the acceleration $\textbf{a}_i$ and first derivative $\dot{\textbf{a}}_i$ acting on a star $i$, a member of a binary or multiple system, caused by an external star $j$ are calculated using
\begin{equation}
 \textbf{a}_{i,\rm pert} = \textbf{a}_{i,j} - \textbf{a}_{\rm{cm},j},
\end{equation}
\begin{equation}
 \dot{\textbf{a}}_{i,\rm pert} = \dot{\textbf{a}}_{i,j} - \dot{\textbf{a}}_{\rm{cm},j},
\end{equation}
where $\textbf{a}_{\rm{cm},j}$ is the contribution of particle $j$ to the acceleration of the center of mass, $\dot{\textbf{a}}_{\rm{cm},j}$ is the contribution of particle $j$ to the first derivative of the acceleration of the center of mass, and $\textbf{a}_{i,j}$ and $\dot{\textbf{a}}_{i,j}$ are the direct contributions of particle $j$ to the total acceleration and jerk of particle $i$, respectively.

Identifying the perturbers of a binary or multiple sub-system is done by using the neighbor-module of GRAPE-6. During the force calculation for an ``virtual'' star $i$, a distance
\begin{equation}
 R_{{\rm p}_i} = \gamma_{\rm pert}^{1/3} \Big( \frac{2m_{\rm g}}{m_i} \Big)^{1/3} d
\end{equation}
is sent to the neighbor-module of GRAPE-6. Where $m_{\rm g}$ is the mass of the most massive star in the system, $m_i$ the mass of the ``virtual'' star, i.e., the total mass of the binary or multiple subsystem, and $d$ the size of the subsystem i.e. the maximum distance between its members. To avoid the case of overflow of the neighbor lists of GRAPE-6, each ``virtual'' star is sent alone to GRAPE-6, after cleaning the previous neighbor lists from GRAPE-6 memory. To avoid overflow and limit the number of perturbers per binary system in systems with tens of thousands of stars, $R_{{\rm p}_i}$ is also modified to ensure that $\sim 100$ neighbors are returned. This is achieved by preventing $R_{{\rm p}_i}$ becoming greater than $2.5 \times r_{\rm 6}$, where $r_{\rm 6}$ is the distance of the sixth nearest neighbor of star $i$, used to find the core radius of the cluster, as we see below, and $2.5 \times r_{\rm 6}$ is an estimate of the distance of the $100th$ nearest neighbor of star $i$, assuming that the density around star $i$ is constant and unchanged from the time of the last calculation of $r_{\rm 6}$. This means that the calculation of the core radius should be repeated relatively often, to estimate the true number density around each star at any time. This calculation is usually iterated a few times every crossing time, which is defined to be
\begin{equation}
 t_{\rm cr} \simeq 2 \sqrt{2} \Big( \frac{r_{\rm v}^3}{GN\bar{m}} \Big)^{1/2},
\end{equation}
as described in Appendix A. With the above restrictions, the GRAPE-6 memory does not overflow and GRAPE-6 returns the identities of the ``neighbors'' of star $i$, which are the perturbers of the corresponding binary system. Nevertheless, if an overflow occurs, then the perturbers are found after searching among all the stars of the system and the result is used to avoid future overflows.

Figure \ref{fig:fig5} shows schematically the distribution of perturbers and non-perturbers for a close binary system.

We note here a couple of definitions used regularly in the remainter of the paper. In all simulations we refer to as escapers, the star clusters that are assumed to be isolated i.e., for which no external galactic field exists. Stars that are moving outwards and reach a distance \textbf{two} times the initial size of the cluster, are considered as escapers from the system. When a star escapes, the energy of the system is reinitialized. In addition we refer to as hard binary a close binary system whose energy per unit mass $h<-1$ or whose semimajor axis $a <R_h(m_i+m_j)/2$, where $m_i$ and $m_j$ the masses of its members and $R_h$ its half-mass radius. In an equal mass system, the limit for $a$ becomes $a<R_{\rm h}/N$, where $N$ is the number of stars. 

\section{Results and performance}

\begin{figure*}[th]
\centering
\includegraphics[width=18cm,height=15cm]{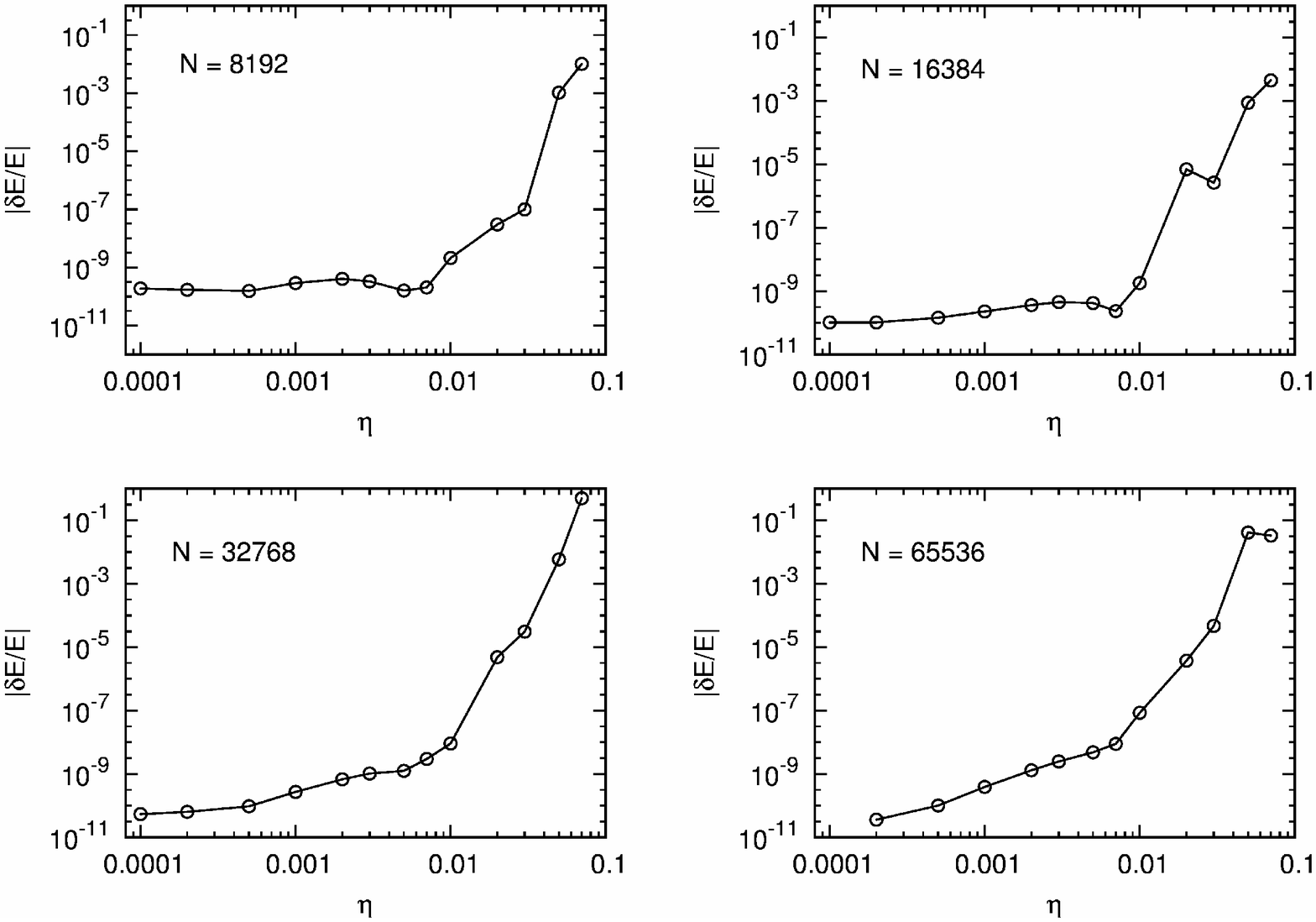}
\caption{Cumulative relative energy error as a function of $\eta$ for a Plummer model with $N=8192$ (top left), $N=16384$ (top right), $N=32768$ (bottom left), and $N=65536$ (bottom right). The integrations ended at $t = 1$ N-body units.}
\label{fig:fig7}
\end{figure*}

\begin{figure*}[t]
\begin{minipage}[t]{0.5\linewidth}
\centering
\includegraphics[height=9cm, width=1\linewidth]{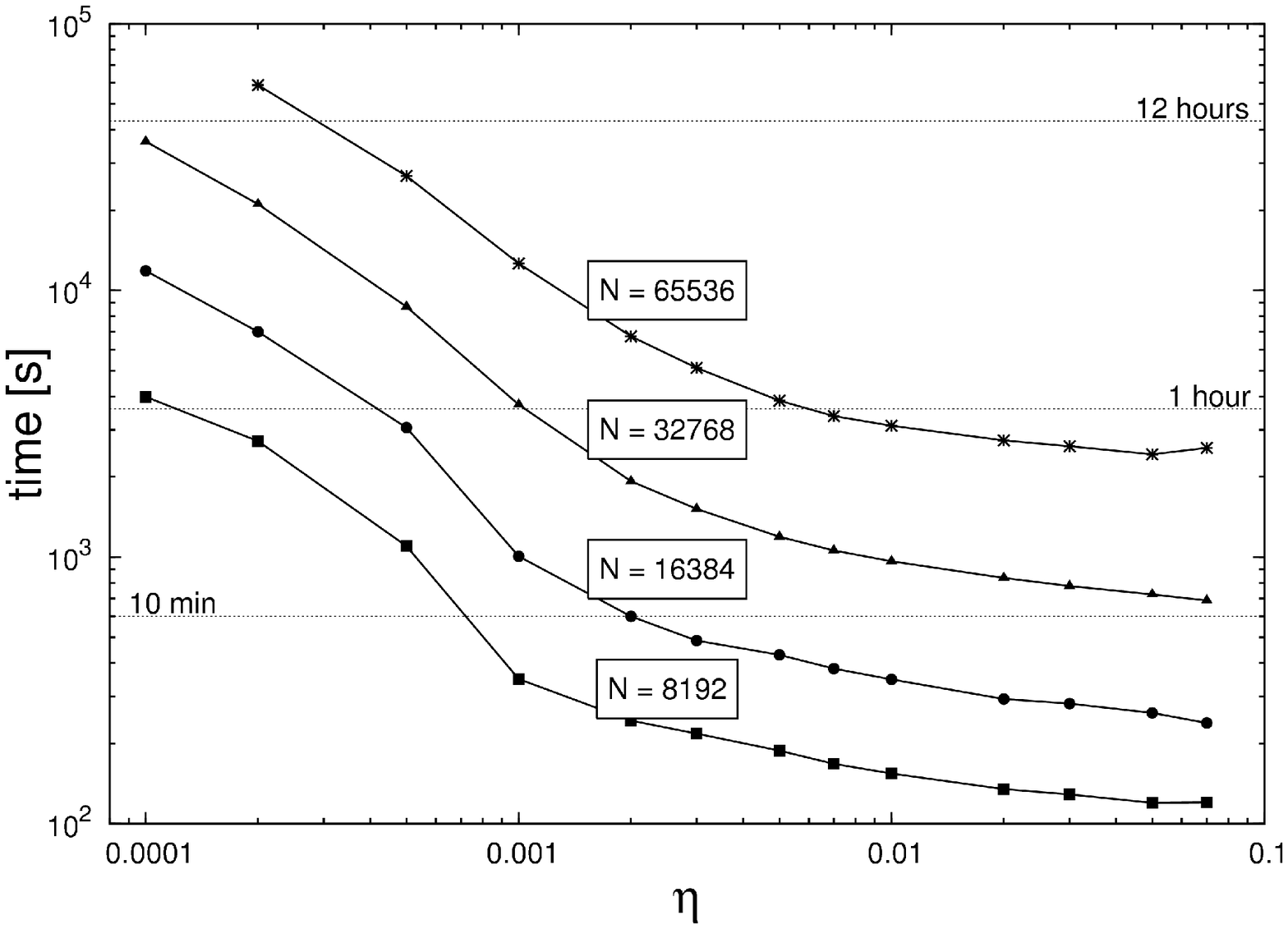}
\caption{Wall-clock elapsed time as a function of $N$ and $\eta$ for the simulations of Figure \ref{fig:fig7}. All simulations ended at $t=1$ in N-body units}
\label{fig:fig8}
\end{minipage}
\hspace{0.2cm}
\begin{minipage}[t]{0.5\linewidth}
\centering
\includegraphics[height=9cm, width=1\linewidth]{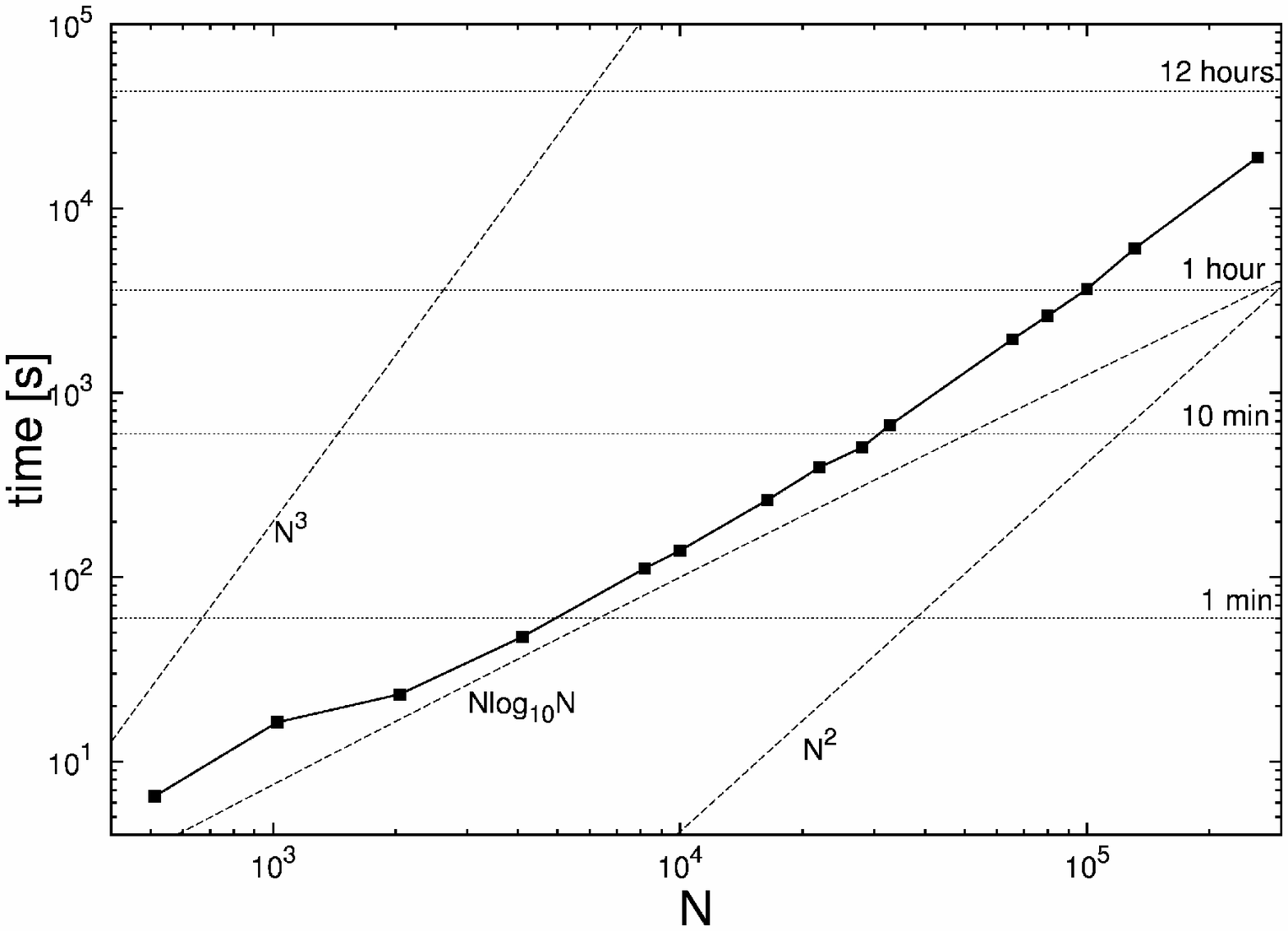}
\caption{Wall-clock elapsed time as a function of the number of stars $N$. All simulations ended at $t=1$ in N-body units. The accuracy parameter $\eta$ had the same value $0.01$ in all simulations. The slopes of $O(N^3)$, $O(N^2)$, and $O(N\log N)$ are  also shown for comparison.}
\label{fig:fig9}
\end{minipage}
\end{figure*}

We measure the accuracy and performance of \texttt{Myriad} by using it to evolve different systems in time. The most simple tests are those where an equal-mass Plummer sphere \citep{Plummer} is used as an initial configuration. We evolve systems with different numbers of stars and measure the accuracy and the speed of \texttt{Myriad}. As a measure of the accuracy of the code, we use the relative error in the total energy, after evolving the systems for one N-body time unit. The speed of the code is measured by the CPU time required for the simulations. We also performed simulations of systems with equal-mass stars distributed in a Plummer sphere and systems with different initial mass functions distributed in either a Plummer sphere or according to a King density profile \citep{King}, up to core collapse. In these simulations, we measure the time of core collapse and compare the results with other codes and with results found in the literature. 

In all simulations, the stars are considered as interacting point particles. The real dimensions of a star are taken into account only when it collides with another star. Stellar evolution, stellar mass-loss, and binary evolution are also not included. If close black-hole binaries are formed, they evolve using purely Newtonian dynamics, without any post-Newtonian terms in the equations. Finally, the effects of the Galactic tidal forces on the star clusters are ignored i.e., star clusters are assumed to be isolated. Initial data in all simulations were provided by the \texttt{Starlab} software environment.

The host computer used for these tests is an AMD Athlon(tm) 64 Processor 3500+ operating at 2.2 GHz with 2GB of RAM, connected to a 32-chip GRAPE-6 board and running Fedore Core 1 GNU/Linux with the 2.4.22 kernel. 

\subsection{Accuracy}

There are many different sources of errors in an N-body code. Two of them, which can be easily controlled, are the choice of integration scheme and its order. Here, as mentioned earlier, the integrator is $4^{th}$ order. If the time step were constant and common for all stars, then the errors would scale as $O(dt^4)$. In codes that use block time steps, such as \texttt{Myriad}, the parameter that controls the error is the accuracy parameter $\eta$ is used in the time-step criterion of Eq. (\ref{Aarseth}). Another source of errors, which is clear when small values of $\eta$ are chosen, is the accuracy of the hardware, both CPU and GRAPE-6. \texttt{Myriad} uses double precision for all calculations on the CPU, while the accuracy of the calculation of the accelerations and derivatives is controlled by GRAPE-6 \citep{Grape6Accuracy}.

We performed a series of experiments using \texttt{Myriad} to evolve equal-mass Plummer models, with different particle numbers $N$ from $t=0$ to $t=1$ N-body time units. For all the integrations, we recorded the cumulative relative energy error ($\Delta E/E = (E_{t=1} - E_{t=0})/E_{0}$) and studied its dependence on the accuracy parameter $\eta$ and particle number $N$. For this study, we varied the accuracy parameter $\eta$ from $10^{-4}$ to $0.2$ and the particle number from $8192(\equiv8K$) to $65536(\equiv64K)$, where $1K \equiv 1024$. The results are shown in Figure \ref{fig:fig7}.

For small values of $\eta$ ($\eta \leq 5\times 10^{-3}$), the relative energy error is almost constant. This is because for those values of $\eta$ the integration error is small and the total error is dominated by the hardware precision. For values of the accuracy parameter $\eta \geq 5\times 10^{-3}$, the error increases with $\eta$ as expected. When $\eta \geq 0.1 $, the errors are too large, so these choices of accuracy parameter are inappropriate for simulations. The typical choices for the accuracy parameter are ( $0.001\leq\eta \leq 0.01 $) and the relative energy error between these limits is $\Delta E/E_0 \leq 10^{-9}$ for $N\leq 16K$ and  $\Delta E/E_0 \leq 10^{-7}$ for greater values of the particle number $N$.

The dependence of the error on the particle number $N$ is weak and evident only at the lower values of $\eta$. For greater values of $N$, the relative energy error saturates to slightly smaller limits.

\subsection{Performance}

The elapsed time $T$ for a simulation depends on the accuracy parameter $\eta$ and particle number $N$. The elapsed time (wall-clock time) as a function of $\eta$ and $N$ for the simulations of Figure \ref{fig:fig7} is shown in Figure \ref{fig:fig8}. As expected, $T$ grows with smaller $\eta$ and greater $N$. We note that smaller $\eta$ indicates that shorter time steps are used on average. 

To study the speed of \texttt{Myriad}, we performed a set of integrations of equal-mass Plummer models with different particle numbers $N$ from $t=0$ to $t=1$ N-body time units. For all simulations, we recorded the wall-clock time as a function of $N$ and the results are shown in Figure \ref{fig:fig9}, where $N$ is chosen to be between $512(=2^9)$ and $262\,144(=2^{18})$. The accuracy parameter used for all simulations was $\eta = 0.01$. In the same figure, the slopes of $N^2$, $N^3$, and $N\log_{10}N$ are shown for comparison.

The complete set of calculations inside \texttt{Myriad} has a time complexity of $O(N\log N)$. There is only one calculation that has a time complexity $O(mN)$, where $1\leq m \leq N$ is the average number of neighbors per particle returned by GRAPE-6 or, if GRAPE-6 returns an overflow, calculated by the CPU. This is the calculation of the core radius of the star cluster. This calculation is repeated every diagnostic time step $dt_{\rm diag}$. For the experiments of Figure \ref{fig:fig9}, this calculation took place only once, before $t=1$. As the number of stars increases, the calculation of the core radius takes a comparable amount of time to the integration time. This is why initially the slope in Figure \ref{fig:fig9} is similar to the slope of $N\log N$, but as $N$ increases, it tends to more closely resemble the slope of $N^2$. We note that all stars at $t=0$ begin with the shortest time step $Dt_{\rm min}$ and then, as time passes by, they are slowly distributed in all the available time blocks. Because of this, the simulation initially operates more slowly. We note that primordial binary systems were not included in all runs, while no close binary systems were dynamically formed until $t=1$ N-body time units. At later times, dynamically formed binary systems would force the simulation to run slower.

\subsection{Binaries and multiples}

\begin{figure*}[t]
\begin{minipage}[t]{0.5\linewidth}
\centering
\includegraphics[width=1\linewidth,height=93 mm]{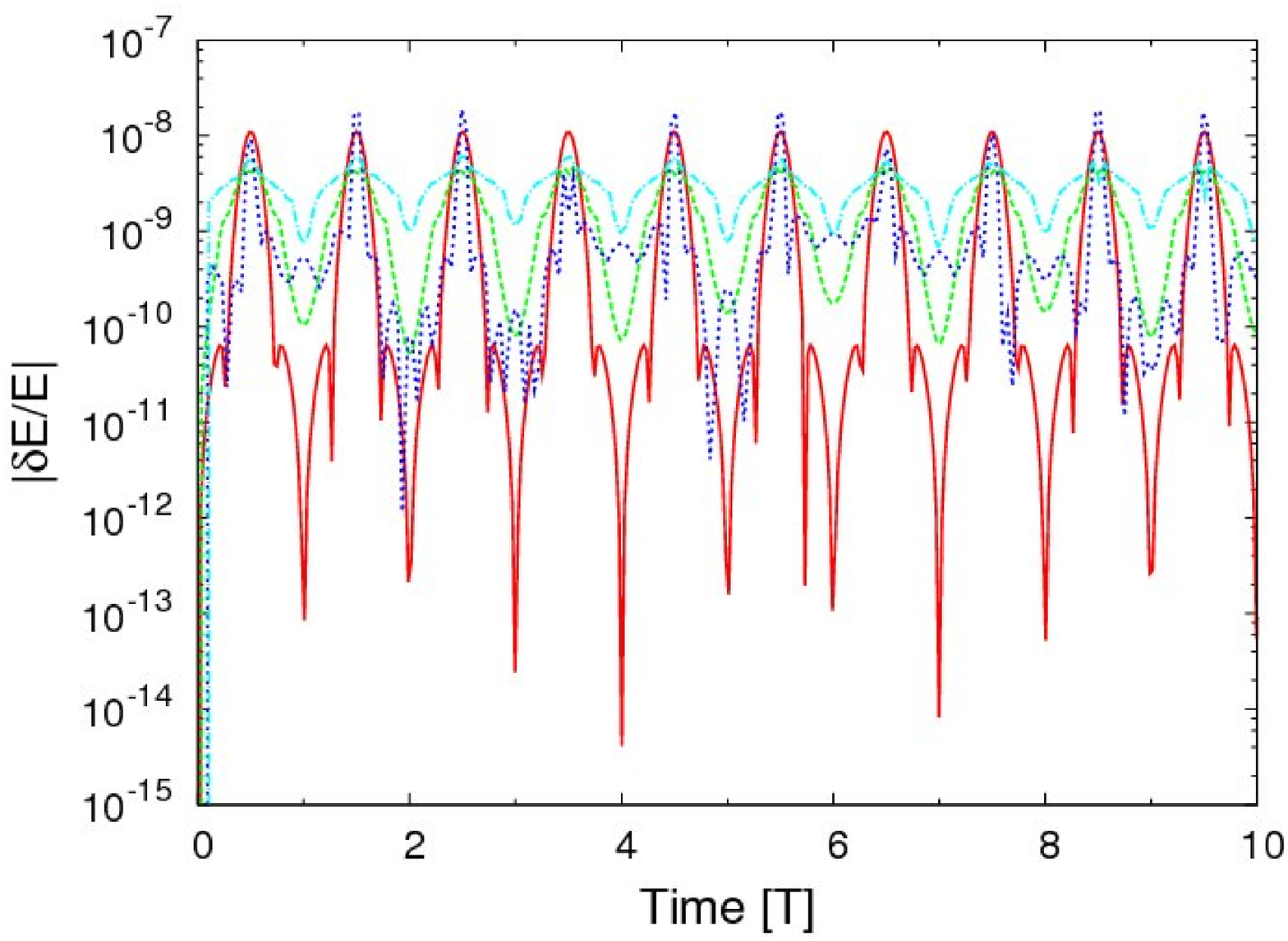}
\caption{Cumulative energy error ($\Delta E/E_{0} = (E_{\rm t} - E_{0})/E_{0}$) as a function of time for the first 10 periods for the simulations of equal-mass binary systems with 4 different eccentricities. The eccentricities of the binaries are $e=0.19$ (red line), $e=0.51$ (green dashed line), $e=0.75$ (blue dotted line), and $e=0.91$ (cyan dashed-dotted line).}
\label{fig:fig_b1}
\end{minipage}
\hspace{0.2cm}
\begin{minipage}[t]{0.5\linewidth}
\centering
\includegraphics[height=93mm, width=1\linewidth]{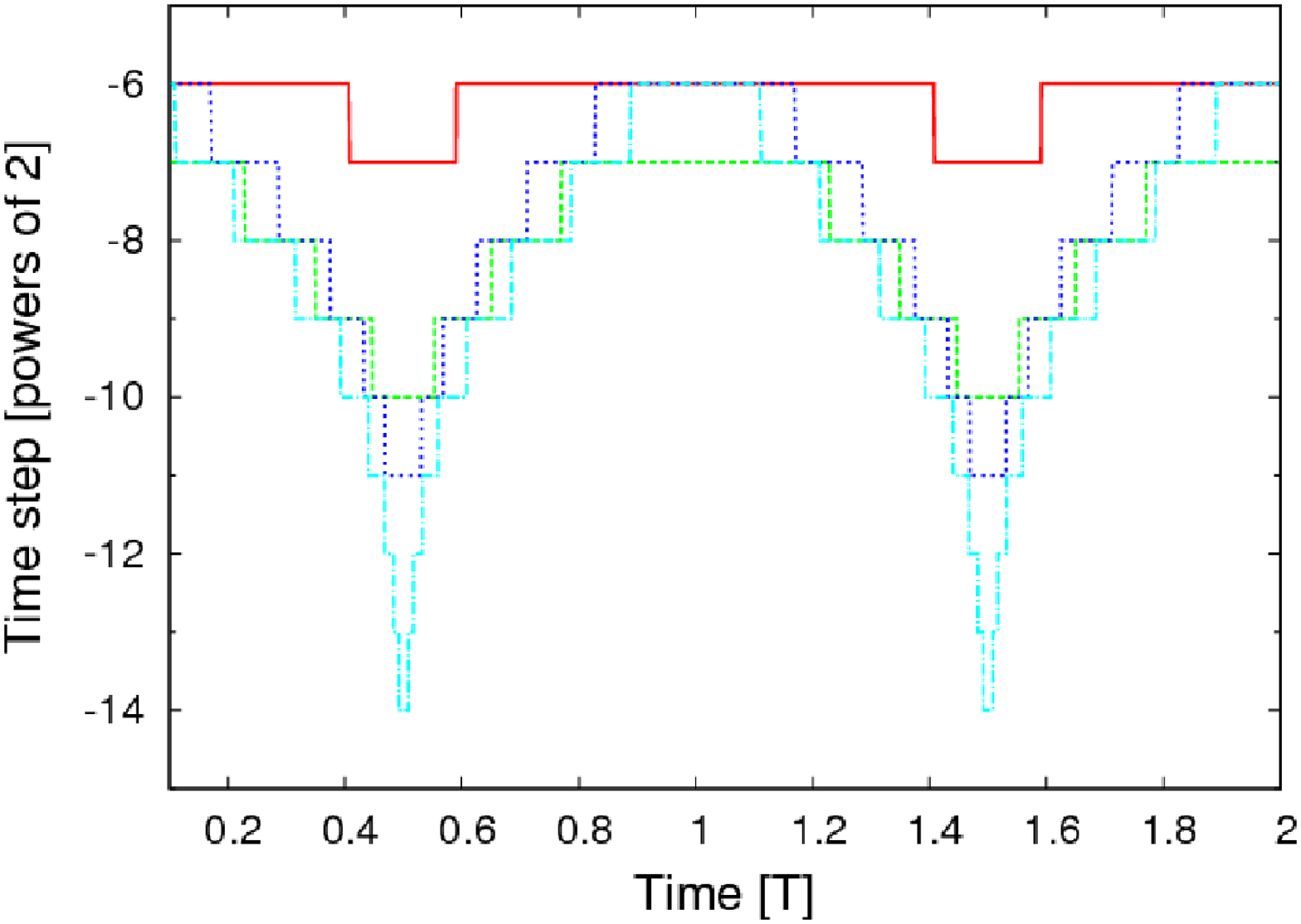}
\caption{Variation in the time step used for the evolution of the binary systems of Figure \ref{fig:fig_b1} during 2 periods of each binary. The time step parameter $\eta_{\rm b}$ is chosen so that the error in all the binaries is of the same order.}
\label{fig:fig_b2}
\end{minipage}
\end{figure*}

\begin{figure*}[!th]
\begin{minipage}{1\linewidth}
\centering
\includegraphics[height=110 mm, width=0.9\linewidth]{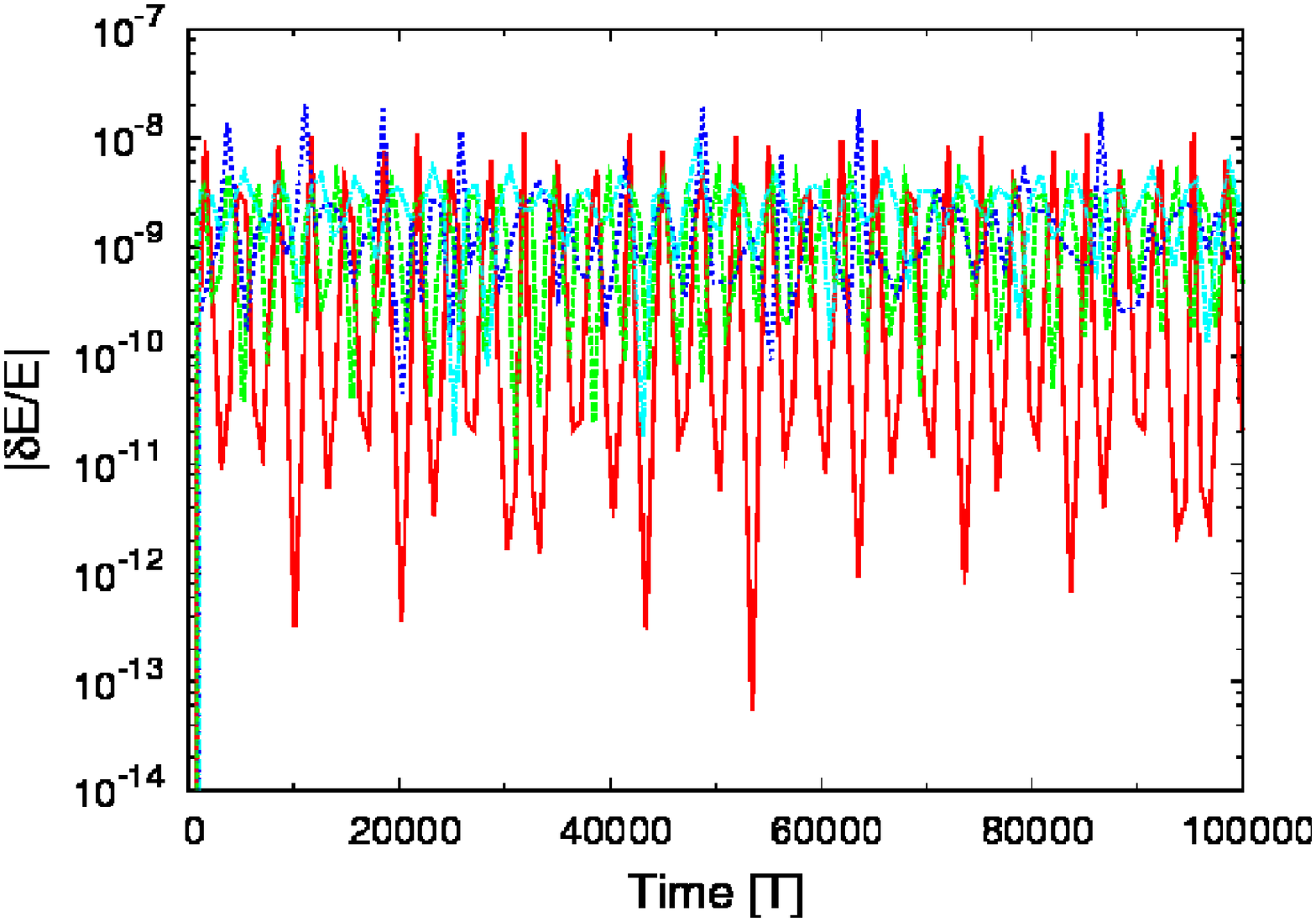}
\caption{Same as Figure \ref{fig:fig_b1}, but for the duration of $10^5$ periods for each binary.}
\label{fig:fig_b3}
\end{minipage}
\end{figure*}

\begin{figure*}[t]
\begin{minipage}[t]{0.5\linewidth}
\centering
\includegraphics[width=1\linewidth,height=9.3cm]{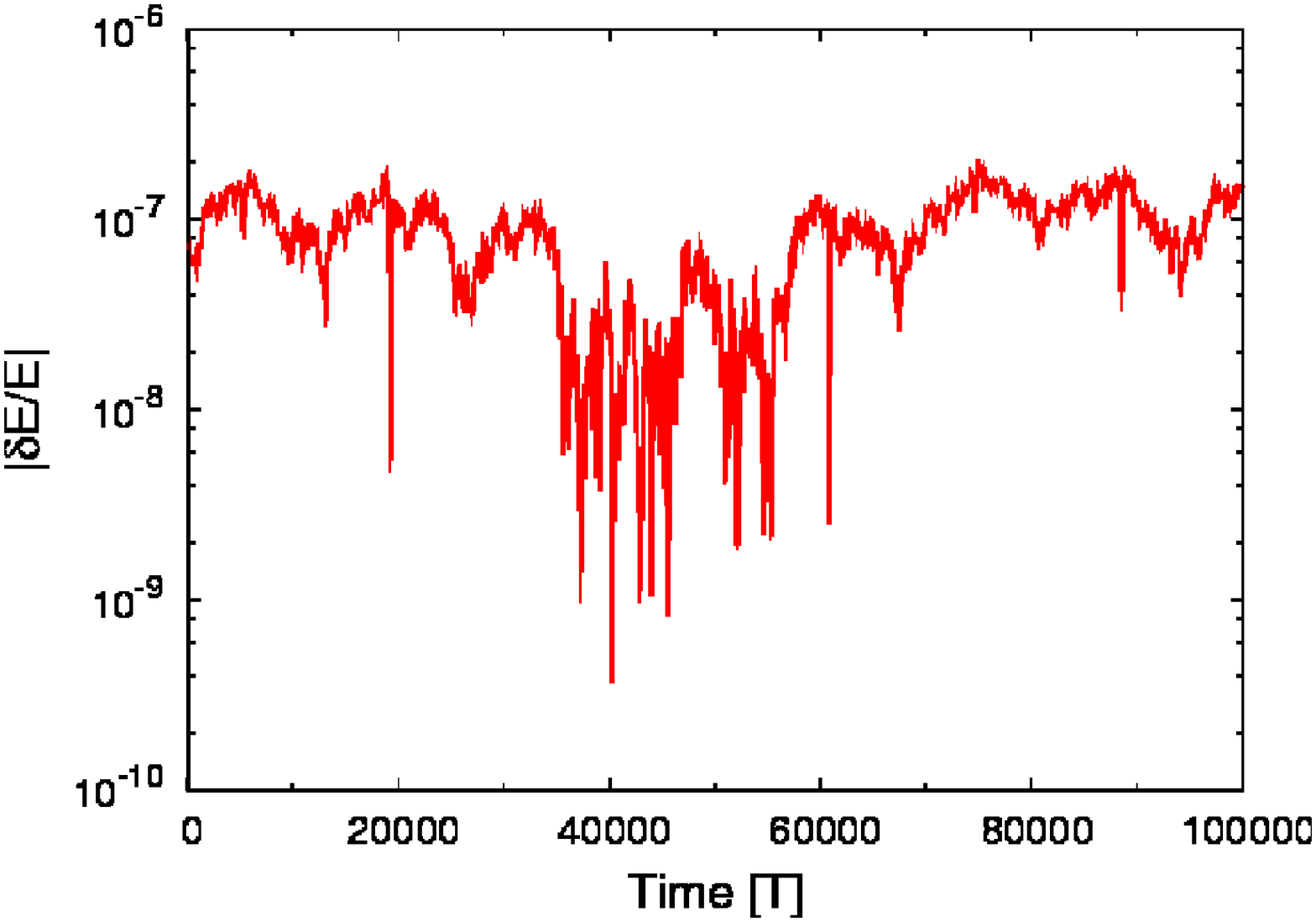}
\caption{Cumulative energy error ($\Delta E/E_{0} = (E_{\rm t} - E_{0})/E_{0}$) as a function of time for $10^5$ periods for the simulation of a binary system with eccentricity $e=0.89$ and mass ratio $m_{heavy}/m_{light} = 15$.}
\label{fig:fig_b4}
\end{minipage}
\hspace{0.2cm}
\begin{minipage}[t]{0.5\linewidth}
\centering
\includegraphics[height=9.3cm, width=1\linewidth]{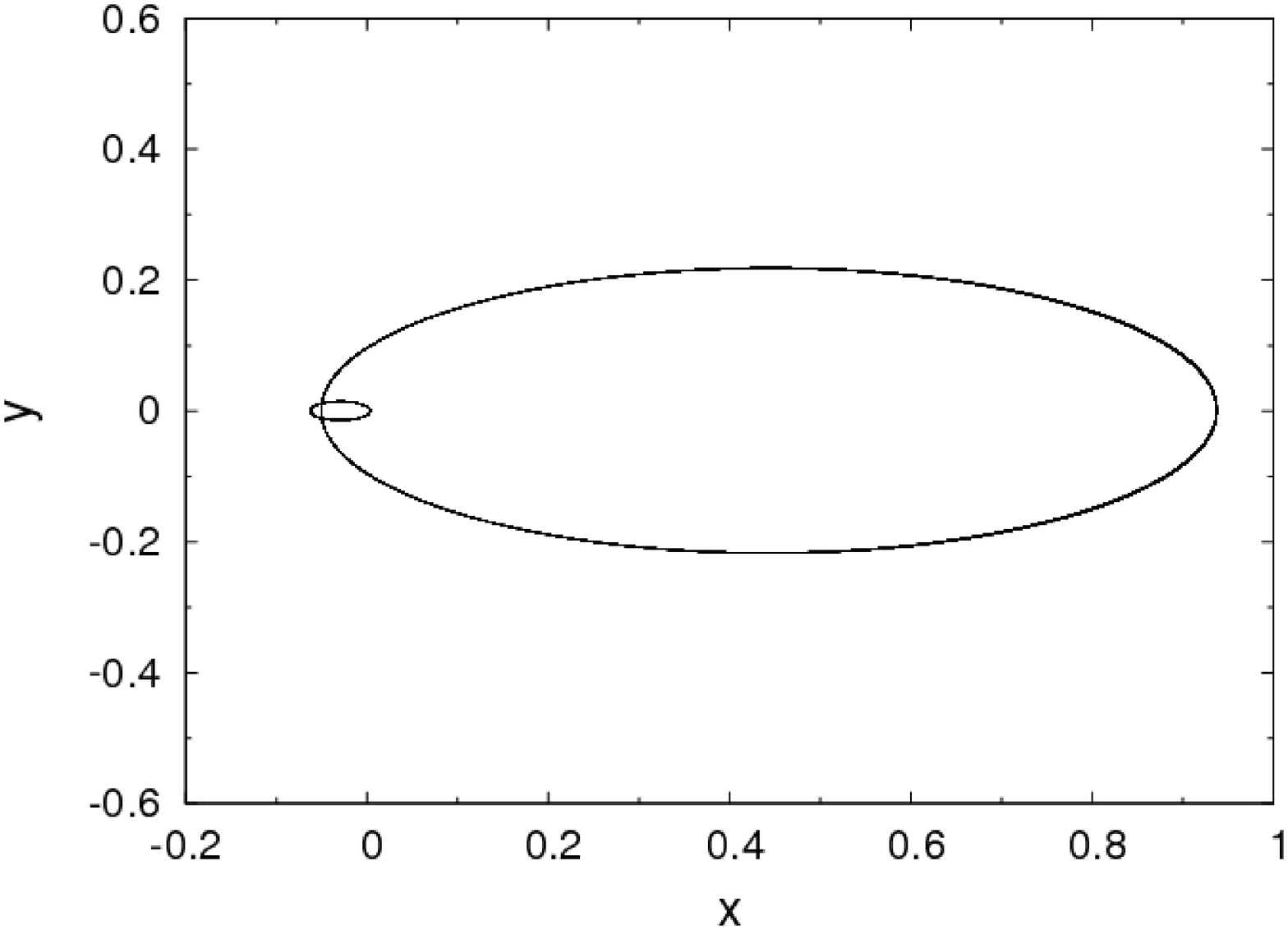}
\centering
\caption{The trajectory of the two stars in the simulation of Figure \ref{fig:fig_b4}. The two stars are orbiting around their common center of mass, which is located at $O(0,0)$.}
\label{fig:fig_b5}
\end{minipage}
\end{figure*}

\begin{figure*}[!th]
\begin{minipage}{0.5\linewidth}
\centering
\includegraphics[width=1\linewidth,height=109.5 mm]{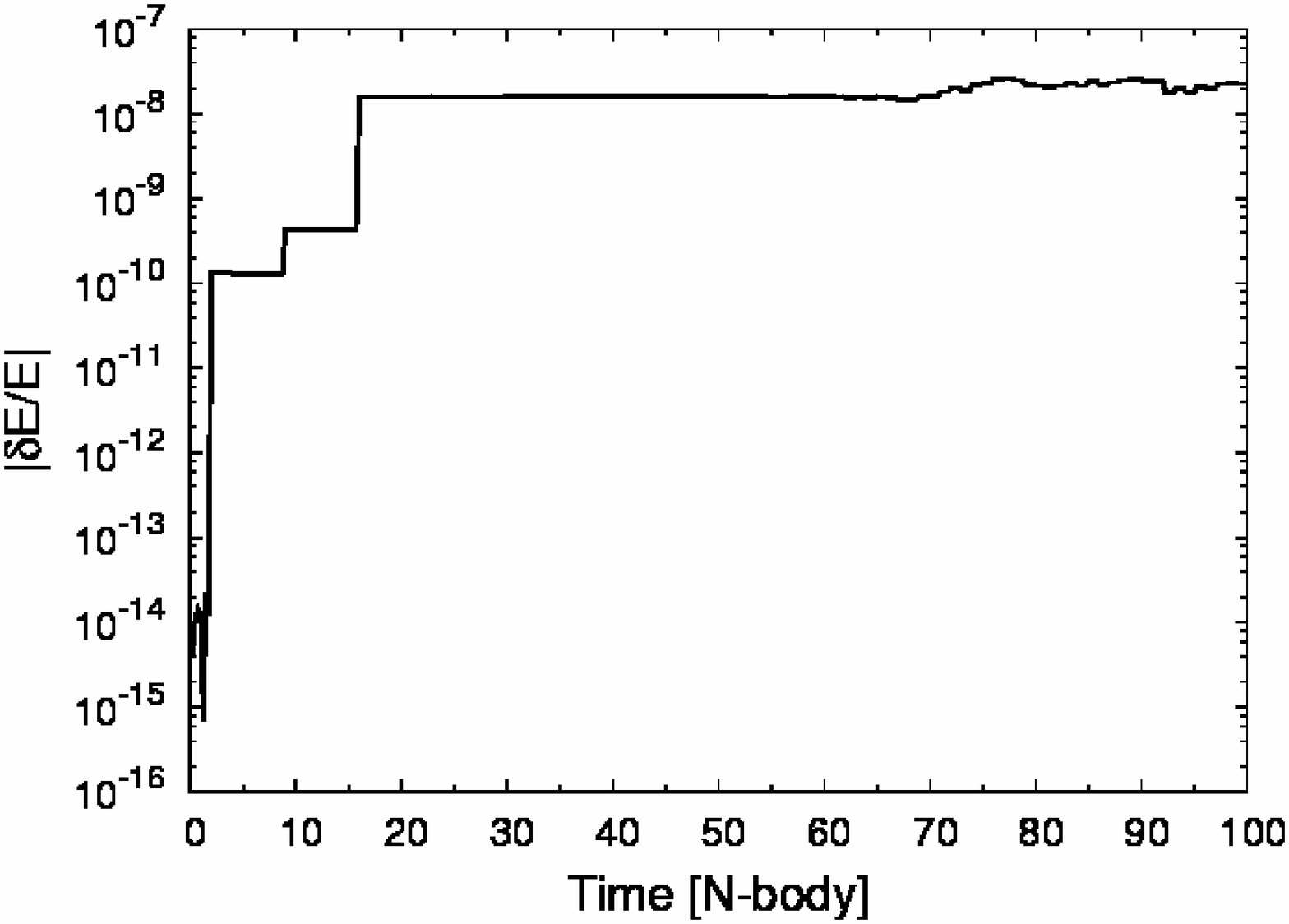}
\caption{Cumulative energy error ($\Delta E/E_{0} = (E_{\rm t} - E_{0})/E_{0}$) as a function of time for the Pythagorean three-body system.}
\label{fig:fig_b6}
\end{minipage}
\hspace{0.1cm}
\begin{minipage}{0.5\linewidth}
\centering
\includegraphics[height=111 mm, width=1\linewidth]{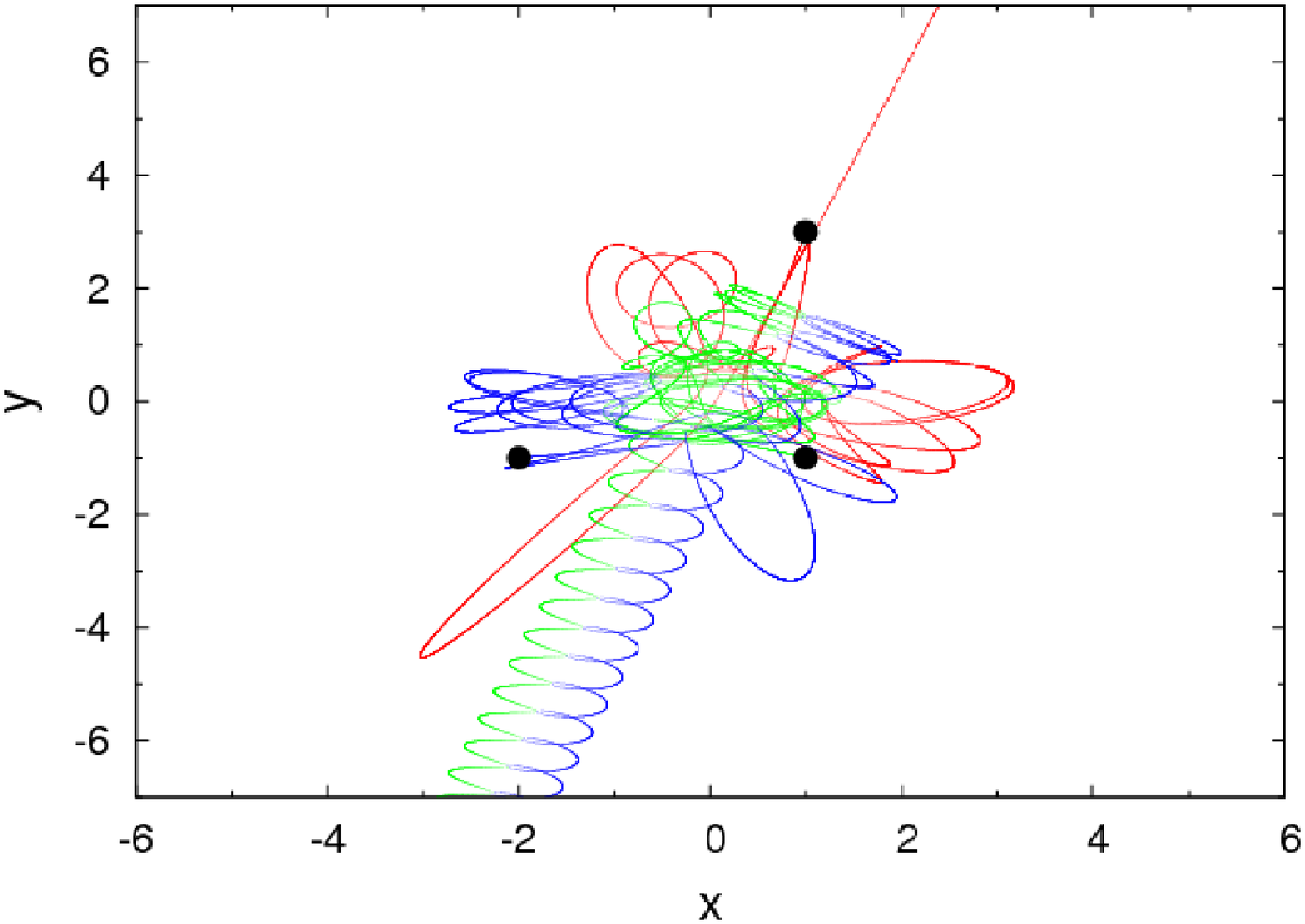}
\caption{Trajectories of the particles in the simulation of the Pythagorean three-body system. The initial positions of the particles are marked with black circles.}
\label{fig:fig_b7}
\end{minipage}
\end{figure*}

Before testing \texttt{Myriad} in simulations of clusters up to core collapse, we investigated the accuracy of the the code in handling close binaries and multiples. We recall that the algorithm used in the binary module of \texttt{Myriad} is the time-symmetric Hermite $4^{th}$ order scheme described in Sect. 2.2.3 and note that the force calculation is done on the CPU.

Figure \ref{fig:fig_b1} shows the cumulative relative error in the energy for the first 10 periods of binaries consisting of equal mass particles and having different eccentricities. The choice of the accuracy parameter $\eta_{\rm b}$ in the time step calculation and the maximum allowed time step are such that for all simulations the initial energy errors are of the same order. The number of time steps per period for the binary with eccentricity $e=0.19$ is 720, for the binary with $e = 0.51$ is 2200, for the binary with $e = 0.75$ is 2200, and for the binary with $e = 0.91$ is 4300. Figure \ref{fig:fig_b2} shows how the time step changes according to our time step criterion, during the first 2 periods of each of the binaries. As shown in the figure, the time step is quantized as a power of 2. Finally, the cumulative energy error for evolving the binaries for $10^5$ periods is shown in Figure \ref{fig:fig_b3}. For a clearer visual result, we have printed a fraction of the points in the figure. It is obvious that the algorithm used for the dynamical evolution of close binary systems is time-symmetric and preserves the energy since there is no linear growth in the energy errors. The energy error oscillates between a lower and higher value. The lower error is observed during the apastron passage, while the higher value is observed during the periastron passage. We would have the same behavior for the energy error even if we decrease the number of steps per period for the simulations. The error would be greater but still have an upper limit and no linear growth with time.

Figure \ref{fig:fig_b4} shows the cumulative relative energy error in the case of a binary with unequal mass members (mass ratio = 15) and eccentricity $e = 0.89$. We followed the simulation for $10^5$ periods. The energy error does not grow linearly with time, but instead follows some kind of random walk between some limits. This result shows that the binary module of \texttt{Myriad} is capable of evolving binaries with high mass ratios and high eccentricities for a long period without significant energy errors. The trajectories of the two particles around their common center of mass for the full simulation of $10^5$ periods are shown in Figure \ref{fig:fig_b5}.

We finally tested the performance of \texttt{Myriad} in handling triple systems by evolving the well known Pythagorean three-body system. This system consists of three different masses initially located at rest at the corners of a right triangle whose sides have lengths 3, 4, and 5. Each of the three masses is located in such a way that its mass is equal to the length of the opposite side. Mass $m_{\rm 1} = 3$ is opposite the side of length 3, $m_{\rm 2} = 4$ is opposite the side of length 4 and $m_{\rm 3} = 5$ is opposite the side of length 5. The three masses interact strongly with each other, and their trajectories after some time are shown in Figure \ref{fig:fig_b7}. The final result is a hard binary consisting of particles with masses $m_{\rm 3} = 5$ and $m_{\rm 2} = 4$ and an escaping particle. The evolution of the cumulative relative energy error for the simulation is shown in Figure \ref{fig:fig_b6}. We note the two ``jumps'' in the energy error. These ``jumps'' occur when two of the particles come very close to each other and as a result more time steps are needed to perform an accurate integration. We used a constant accuracy parameter $\eta_{\rm b} = 0.0001$. The cumulative relative energy error at $t=100$ N-body units is $2.23776 \times 10^{-8}$.

\subsubsection{Choice of the parameters}

In \texttt{Myriad}, assuming that a good choice for the accuracy parameter $\eta$ is $0.01$ and adjusting the accuracy parameter for binary evolution $\eta_{\rm b}$ so that there is a smooth transition between the binary module and the Hermite integrator, there is another free parameter that balances between speed and accuracy in collisional simulations up to core collapse. This is the number of perturbers for every binary or multiple subsystem controlled by the critical dimmensionless perturbation $\gamma_{\rm pert}$. This parameter is discussed in Sect. 2.2.3 (Eq. (\ref{gamma_pert1})). For the rest of this work, we choose $\gamma_{\rm pert} = 10^{-7}$, which is small enough to ensure that the errors introduced into the simulations by replacing two or more particles that lie close to each other, with their center of mass, are not significant. Smaller values of $\gamma_{\rm pert}$ would slightly increase accuracy, but the speed of the code would be lowered significantly, since the forces of the perturbers on the members of a binary are computed on the CPU and not on the GRAPE-6. Finally, if the value of $\gamma_{\rm pert}$ is responsible for large numbers of perturbers per binary system, this value is increased so that no more than $100$ perturbers per binary are used.

\subsection{Core collapse}

The evolution of a star cluster up to and beyond core collapse is one of the challenges for any N-body code. During core collapse, the density at the center reaches its maximum value, while the core radius, defined by the equation
\begin{equation}
 r_{\rm c} = \sqrt{\frac{\sum_{i} \rho_i^2 | \textbf{r}_i - \textbf{r}_{\rm d}|^2}{\sum_i \rho_i^2}},
\end{equation}
which is described in Appendix A, reaches its minimum value. Figures. \ref{fig:fig14a} and \ref{fig:fig11} clearly show this behavior as a star cluster of $1024$ equal-mass stars evolves beyond core collapse. As the cluster evolves, close binary or multiple systems are formed and interact with single stars or other binary or multiple systems. Those interactions become more frequent and violent as the cluster approaches core collapse, because the density of stars around the center of the cluster reaches its maximum. All of these encounters have to be resolved with an acceptable accuracy. In addition, some stars approach the outer bounds of the system and consequently escape the system.

We tested the behavior of \texttt{Myriad} in evolving star clusters up to core collapse and compared the results with those produced by \texttt{Starlab} or found in the literature. For these tests, we used $3$ different initial configurations. We initially used an equal-mass Plummer model and studied the time of core collapse and the evolution of the error in the total energy. The same simulation with similar parameters was performed using \texttt{Starlab} for comparison. The next test was the evolution of a star cluster with an initial mass function up to core collapse. The stars were again distributed initially according to a Plummer density profile. Finally, we performed several simulations of star clusters with an initial mass function of a broad range of masses, distributed initially according to a King density profile. The result of the latter simulations was a measure of the core collapse time ($t_{\rm {cc}}$) as a function of the half-mass relaxation time ($t_{\rm rlx}$). These results were compared with existing results found in the literature.

\subsubsection{Equal-mass Plummer models}

\begin{figure*}[!t]

\begin{minipage}[t]{0.5\linewidth}
\centering
\includegraphics[height=9cm, width=1\linewidth]{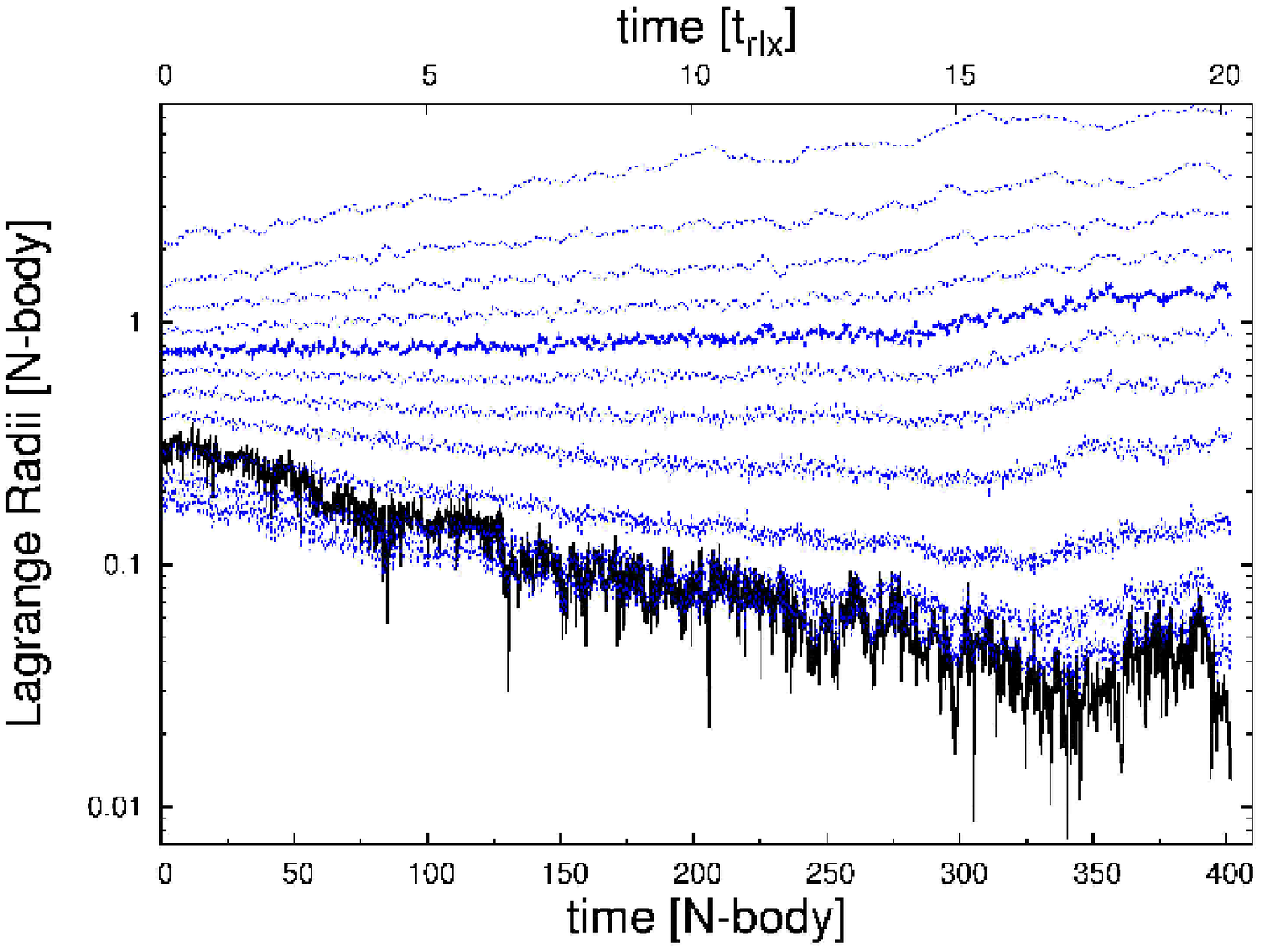}
\caption{Core radius (heavy black line) and Langrangian radii containing $90\%$, $80\%$, $70\%$, $60\%$, $50\%$, $40\%$, $30\%$, $20\%$, $10\%$, $5\%$, and $3\%$ (blue dashed lines from top to bottom) of the total mass. The half mass radius is indicated by a heavier dashed line. Core collapse is reached at $t_{\rm cc} \simeq 340$ N-body time units or $t_{\rm cc} \simeq 17t_{\rm rlx}$  }
\label{fig:fig11}
\end{minipage}
\hspace{0.2cm}
\begin{minipage}[t]{0.5\linewidth}
\centering
\includegraphics[height=9cm, width=1\linewidth]{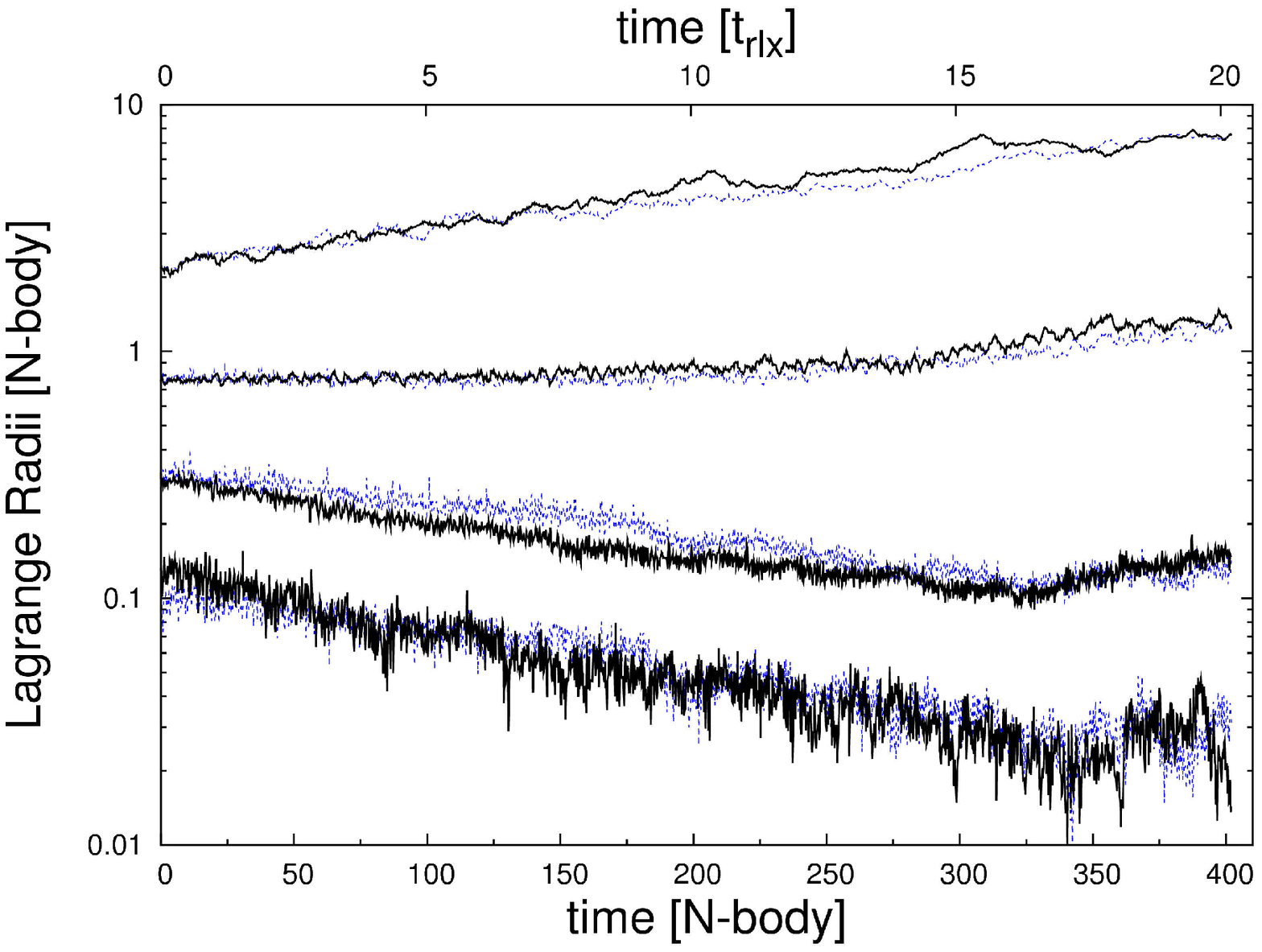}
\caption{Lagrangian radii containing from top to bottom $90\%$, $50\%$, $10\%$, and $1\%$ of the total mass. The continuous lines are the results produced by \texttt{Myriad}, while the blue dashed lines are produced using \texttt{Starlab}.}
\label{fig:fig15a}
\end{minipage}

\begin{minipage}[t]{0.5\linewidth}
\centering
\includegraphics[width=1\linewidth,height=9cm]{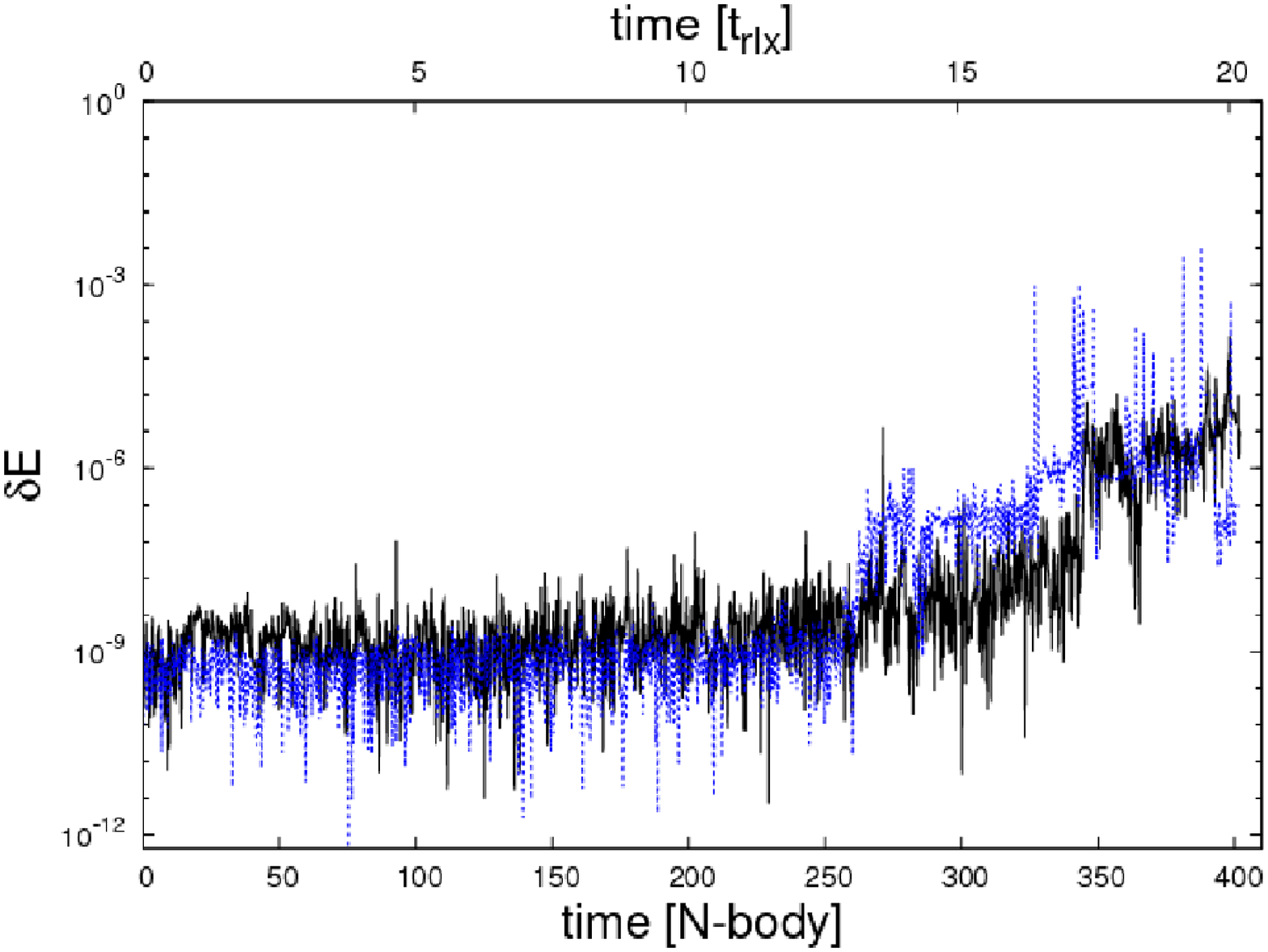}
\caption{Error in the energy per quarter of the time unit ($\delta E = E_{\rm (t)} - E_{\rm{(t-0.25)}}$) as it evolves in time for the simulation of an equal-mass Plummer model of 1024 stars. The simulation ended at $t \simeq 402$ N-body time units. The dashed line is the energy error of \texttt{Starlab} when it was used for evolving the same system. Before core collapse, the  energy errors are identical.}
\label{fig:fig12}
\end{minipage}
\hspace{0.2cm}
\begin{minipage}[t]{0.5\linewidth}
\centering
\includegraphics[width=1\linewidth,height=9cm]{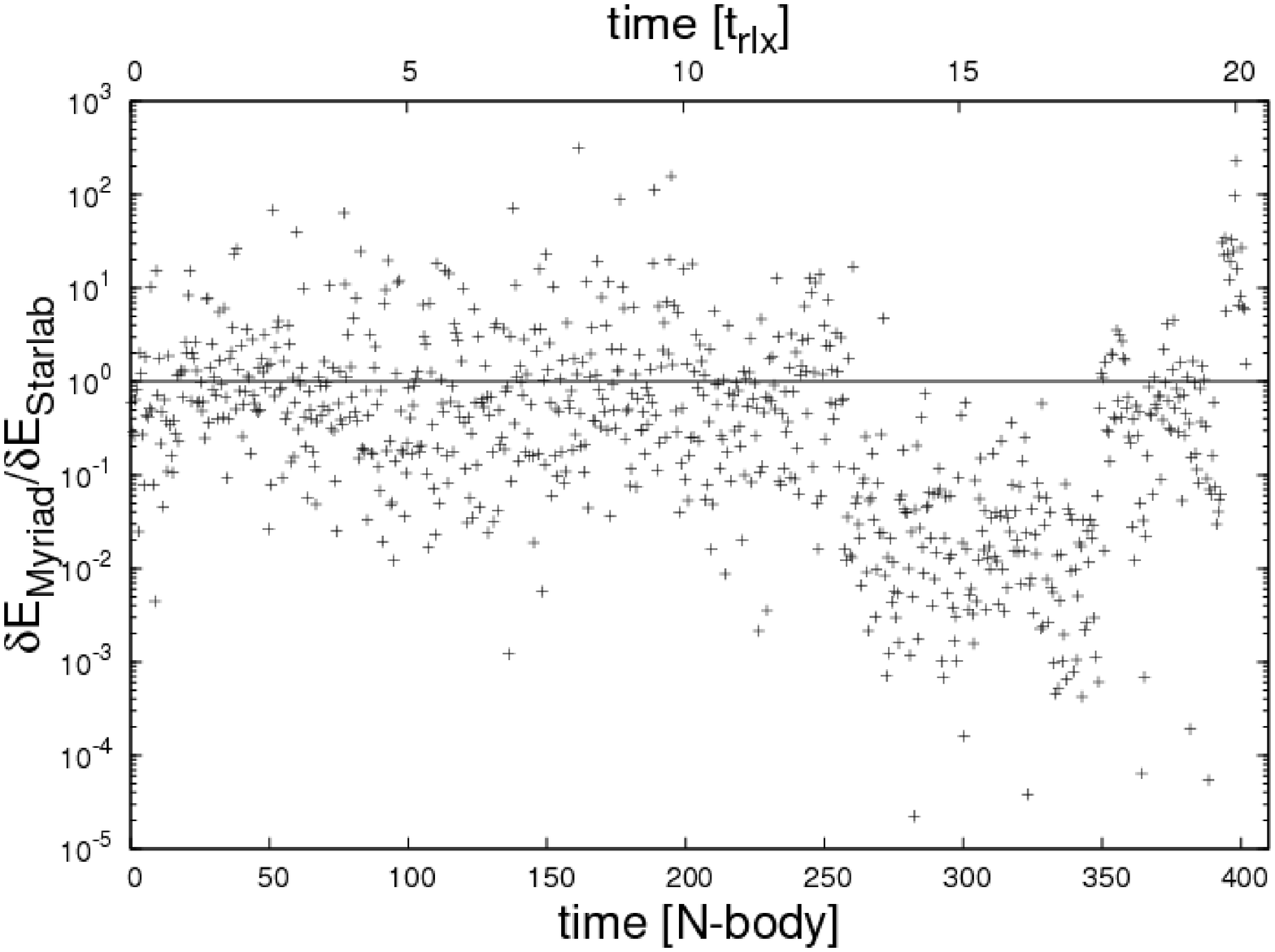}
\caption{Respective energy error ratio \texttt{Myriad}/\texttt{Starlab} as a function of time.}
\label{fig:fig11a}
\end{minipage}
\vspace{2cm}
\end{figure*}

\begin{figure*}[!t]
\begin{minipage}[t]{0.5\linewidth}
\centering
\includegraphics[width=1\linewidth,height=9cm]{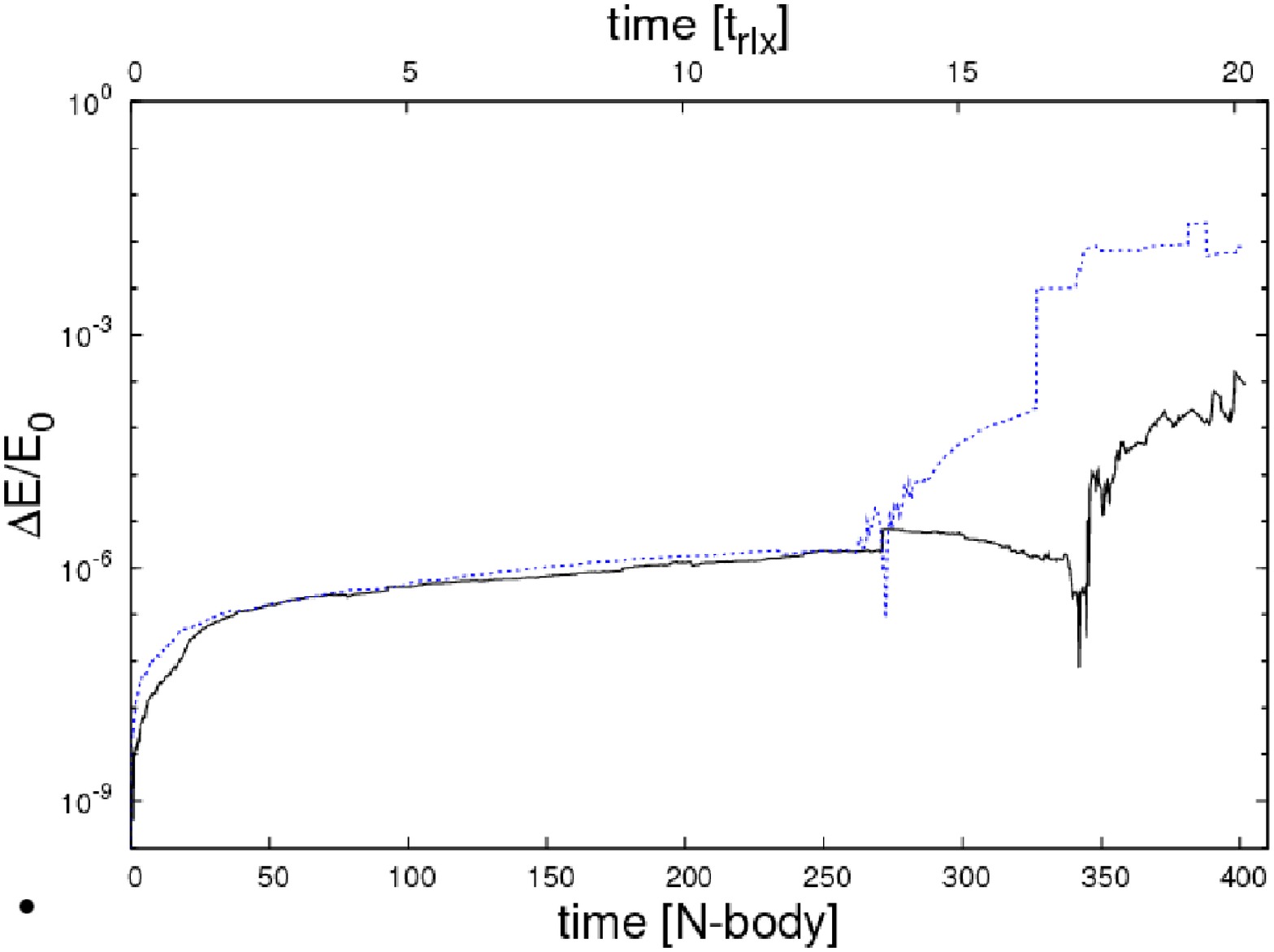}
\caption{Cumulative energy error ($\Delta E/E_{0} = (E_{\rm t} - E_{0})/E_{0}$) as a function of time for the simulation described in Figure \ref{fig:fig12}. The dashed line is the energy error of \texttt{Starlab}, while the heavy black line is the energy error of \texttt{Myriad}.}.
\label{fig:fig12a}
\end{minipage}
\hspace{0.2cm}
\begin{minipage}[t]{0.5\linewidth}
\centering
\includegraphics[width=1\linewidth,height=9cm]{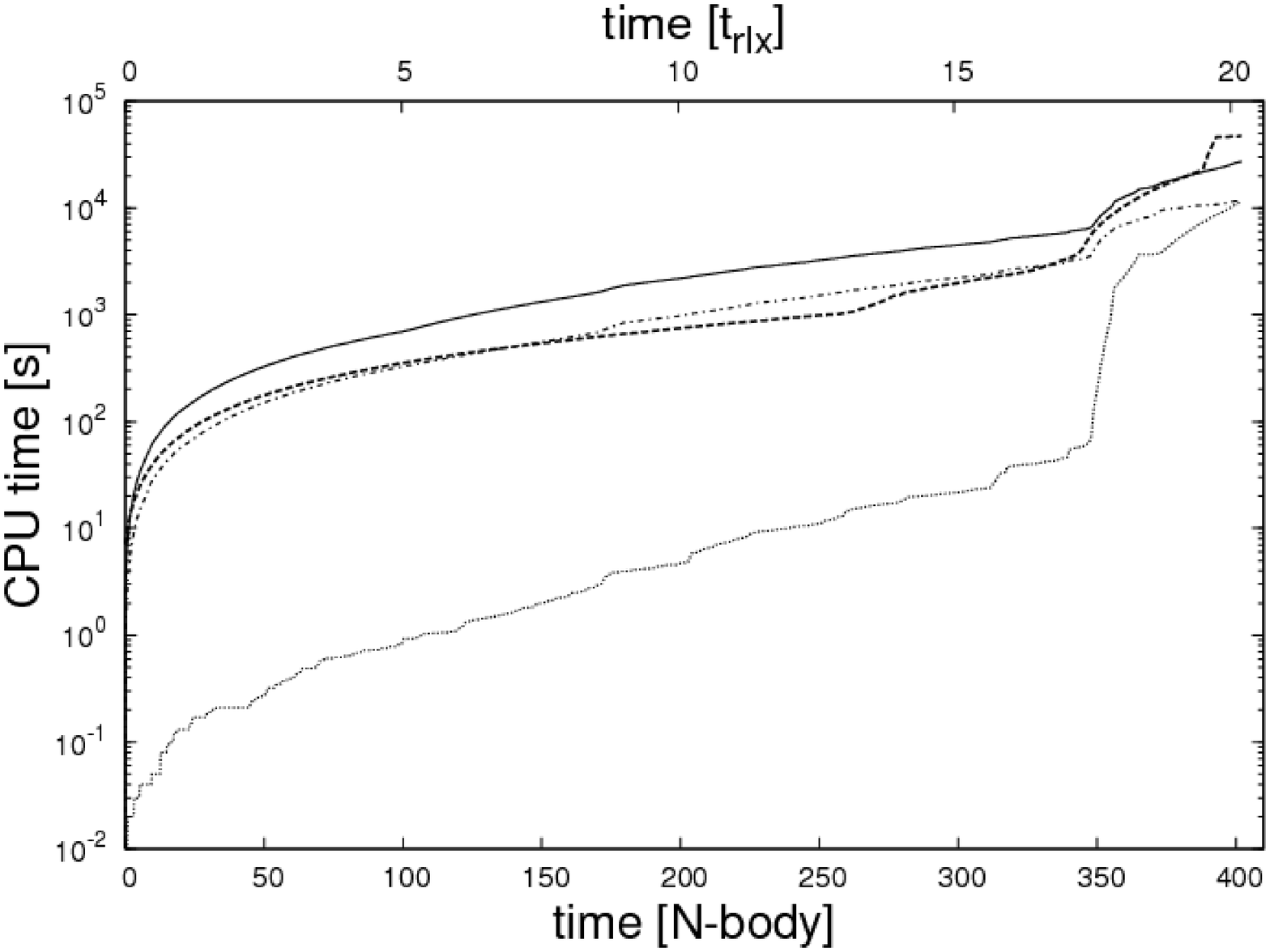}
\caption{CPU time for the simulation as a function of simulation time. The continuous line is the total CPU time of \texttt{Myriad}, while the heavy dashed line is the total CPU time of \texttt{Starlab}. The dashed-dotted line shows the CPU time taken for the force calculation and neighbor list creation done on GRAPE-6. The dotted line is the CPU time required for the evolution of binary systems.}
\label{fig:fig15b}
\end{minipage}

\begin{minipage}[t]{0.9\linewidth}
\centering
\includegraphics[height=10.5cm, width=0.7\linewidth]{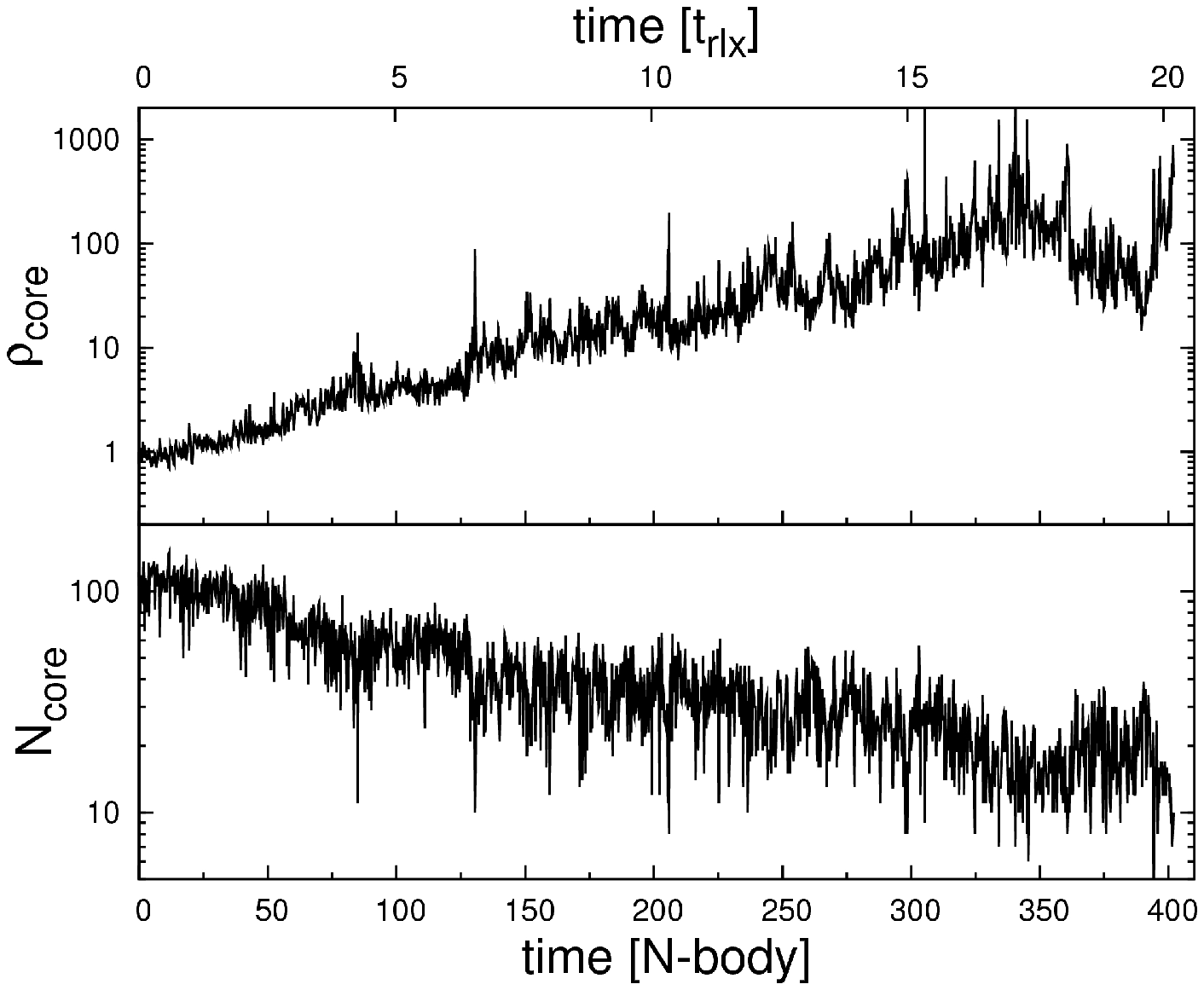}
\caption{Core mass-density (in N-body units) (top) and number of stars in the core (bottom). At the time of core collapse the core density reaches a maximum value}
\label{fig:fig14a}

\end{minipage}
\vspace{1cm}
\end{figure*}

We performed a simulation of an equal-mass Plummer model up to core collapse to test the stability, accuracy, and speed of \texttt{Myriad}. Figure \ref{fig:fig11} shows the evolution of the Lagrangian radii that corresponds to fractions of the total mass of the system and the evolution of the core radius of a system, which was initially an equal-mass Plummer model with $N=1024$ stars. It is obvious that the core radius and the smallest Lagrangian radii reach a minimum at $t_{\rm cc}\simeq340$ N-body time units, when core collapse occurs. From equation
\begin{equation}
 t_{\rm rlx} = 0.138 \Big( \frac{N r_{\rm h}^3 }{G\bar{m}} \Big)^{1/2} \frac{1}{\ln{\Lambda}} ,
\end{equation}
which is described in Appendix A, we find that for this system the initial half-mass relaxation time is $t_{\rm rlx} = 19.92$ N-body time units, so $t_{\rm cc} \simeq 17 t_{\rm rlx}$, which is close to what is expected from Eq. (\ref{cc_1}) and consistent with the results of other codes presented on Table 1 of \cite{Freitag1} and \cite{StarNbodyCompare}, where a detailed comparison between the two major N-body codes \texttt{Starlab} and \texttt{NBODY4} is presented. In Figure \ref{fig:fig15a}, we compare the results of \texttt{Myriad} and  \texttt{Starlab} for the evolution of selected Lagrangian radii of the same system. The small differences in the Lagrangian radii computed by the two codes, may be caused by the difference in the escaper-removal criterion and small differences in the time-step allocation criteria. Because of these differences, the simulation performed with \texttt{Myriad} ended with $983$ stars remaining in the system, while that completed with \texttt{Starlab} contained $973$. Figure \ref{fig:fig12} shows the evolution of the instantaneous energy error $\delta E$ for this simulation. In the same figure, the simulation error of \texttt{Starlab} is presented for comparison. The error is initially small and unaffected by close encounters and binary systems. It grows slowly with time, as expected for the H4 integrator. As the system approaches core collapse, the interactions at the center become more frequent and violent and the energy error is contolled by them. Hard binaries form and their interactions with single stars or other binaries introduce another source of error. After core collapse, the average error becomes smaller, but some peaks, due to strong encounters between hard binaries, continue to appear. 

The comparison between \texttt{Myriad} and \texttt{Starlab} is also shown in Figure \ref{fig:fig11a}, where the respective relative energy error ratio \texttt{Myriad}/\texttt{Starlab} is presented. The error of \texttt{Myriad} at any time step is smaller by 300 times than the error of \texttt{Starlab}, while most of the time the error of \texttt{Starlab} is greater. After $t\sim 250$ and until core collapse, the error of \texttt{Myriad} becomes several times smaller than that of \texttt{Starlab}. After core collapse, both \texttt{Starlab} and \texttt{Myriad} have large instantaneous errors and their ratio scatter from $10^{-5}$ to $300$. Figure \ref{fig:fig12a} shows the cumulative relative energy error for the simulation as it evolves with time. The same error of \texttt{Starlab} again is presented for comparison. It is obvious that before core collapse the two codes show identical behavior and after core collapse the error of \texttt{Myriad} remains smaller than that of \texttt{Starlab}.

The simulation ended at $t \simeq 402$ N-body time units. In Figure \ref{fig:fig15b}, the CPU times of \texttt{Myriad} and \texttt{Starlab} are compared. For \texttt{Myriad}, the total CPU time and the CPU time spent in force calculation and binary evolution are shown. Before core collapse, the CPU time of \texttt{Myriad} is constantly greater than that of \texttt{Starlab}. This is because of the small differences between the two codes in the time step criterion, which on average causes \texttt{Myriad} to store more particles in small time-step blocks. A search for neighbors also occurs every time these particles are updated, which slows down the simulation significantly. The CPU time of the \texttt{Myriad} simulation until core collapse is about $T \simeq 1.7$ hours, while the same time for \texttt{Starlab} is about $T \simeq 1$ hour. After core collapse, the CPU time of \texttt{Starlab} becomes longer because of the formation of hard binaries. Hard binaries lead to an increase in the CPU time of \texttt{Myriad} cause it to finish in about $T \simeq 8$ hours. The simulation done with \texttt{Starlab}, with similar parameters as those used in \texttt{Myriad}, took about $13$ hours, spending most of the time in the post-collapse part. Finally, in Figure \ref{fig:fig14a} we show the evolution of the mass density and the number of stars at the core. As expected, the core density increases with time and reaches a maximum at the time of core collapse. On the other hand, the number of stars at the core drops with time and reaches a minimum value at core collapse.

\subsubsection{Plummer models with a Salpeter initial mass function}

\begin{figure*}[!t]
\begin{minipage}[t]{0.5\linewidth}
\centering
\includegraphics[height=8cm, width=1\linewidth]{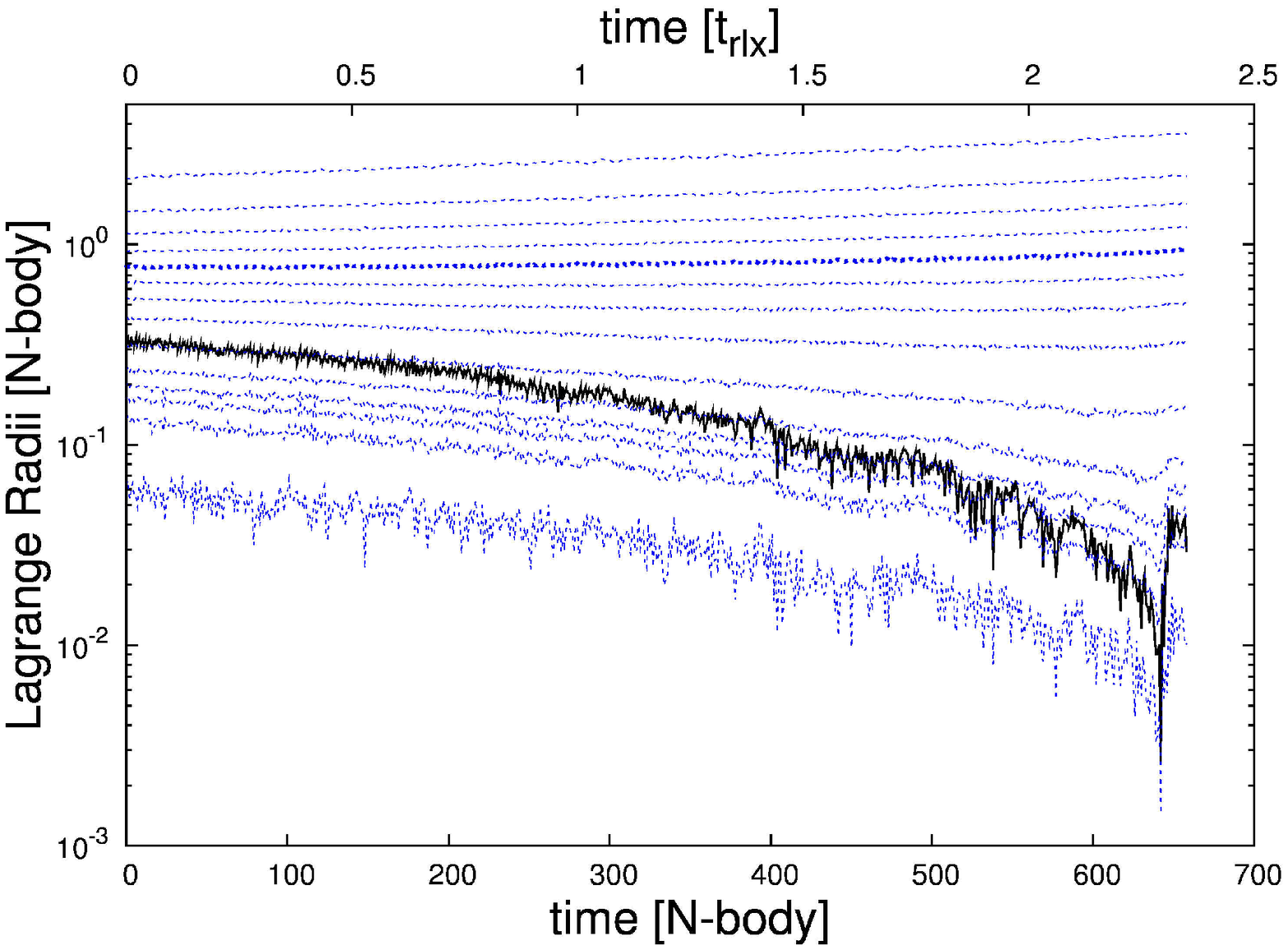}
\caption{Core radius (heavy black line) and Lagrangian radii containing $90\%$, $80\%$,  $70\%$, $60\%$, $50\%$, $40\%$, $30\%$, $20\%$, $10\%$, $5\%$, $3\%$, $2\%$, $1\%$, and $0.1\%$ (blue dashed lines from top to bottom) of the total mass. The half-mass radius is indicated by a heavier blue dashed line. The initial condition was a Plummer model of 16384 stars. A Salpeter mass function with mass limits $m_{\rm lower} = 0.5 M_\odot$ and $m_{\rm upper} = 5 M_\odot$ was chosen as the initial mass function. Core collapse is reached at $t_{\rm cc}\simeq 634$ N-body time units or $t_{\rm cc} \simeq 2.2 t_{\rm rlx}$. The simulation ended at $t \simeq 660$ N-body time units and took $\sim 5$ days}
\label{fig:fig13}
\end{minipage}
\hspace{0.5cm}
\begin{minipage}[t]{0.5\linewidth}
\centering
\includegraphics[height=8cm, width=1\linewidth]{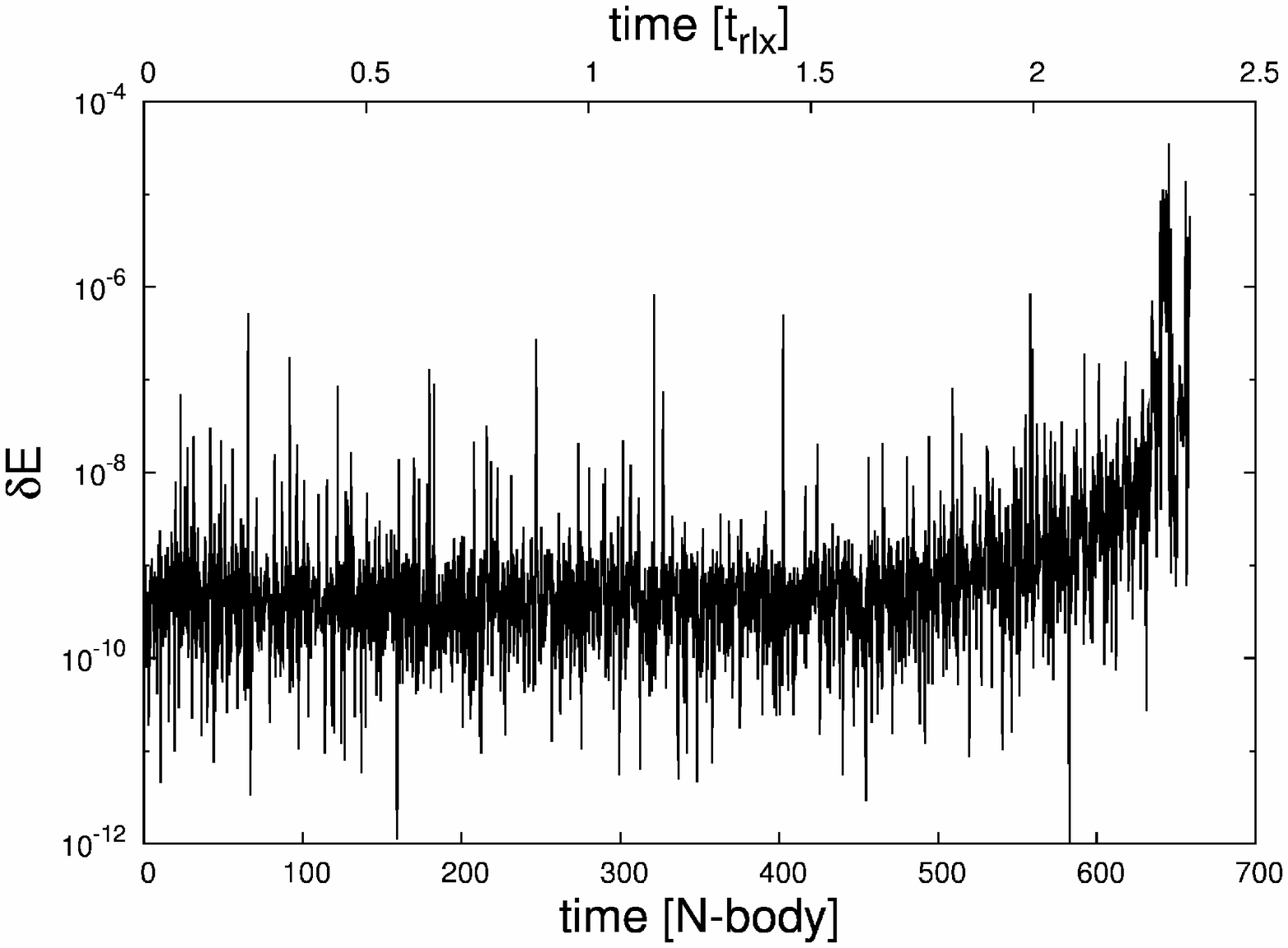}
\caption{Error in the energy per quarter of the time unit ($\delta E = E_{\rm (t)} - E_{\rm {(t-0.25)}}$) as it evolves in time for the simulation of Figure \ref{fig:fig13}}
\label{fig:fig14}
\end{minipage}
\vspace{1cm}
\end{figure*}

A simulation of a Plummer model with $N=16\,384$ stars was performed assuming a Salpeter \citep{Salpeter} initial mass function (IMF) and limiting masses of $m_{\rm lower} = 0.5 M_\odot$ and $m_{\rm upper} = 5 M_\odot$. In Figure \ref{fig:fig13} we show the evolution of the Lagrangian radii and core radius of the system. In Figure \ref{fig:fig14}, we illustrate the evolution of the error in the total energy of the system for this simulation. Core collapse is reached at $t\simeq 2.2 t_{\rm rlx}$, which is much earlier than expected for an equal-mass system, given empirically by Eq. (\ref{cc_1}). This is expected because heavier stars tend to sink to the core faster and because of that core collapse occurs earlier. The time taken for a star of mass $m$ to sink to the center from a circular orbit at $r \gg r_{\rm c}$ is given by \citep{Zwart2003}
\begin{equation}  \label{t_s}
 t_{\rm s} = 3.3 \frac{\bar{m}}{m}t_{\rm rlx},
\end{equation} 
According to Eq. (\ref{t_s}), the most massive stars of the above system would sink to the center in a time of about
$t_{\rm s} \simeq 0.73 t_{\rm rlx}$. Until $t = 2.2 t_{\rm rlx}$, most of the massive stars are located close to the center and this leads to the observed core collapse of the cluster.

\subsubsection{King models with an initial mass function}

\begin{table*}[t]
\centering 
\caption{The core collapse times for different initial models}
\begin{tabular} {c c c c c c}
 \hline \hline
Model & N & runs & $\langle t_{\rm rlx} \rangle$ [Myr] & $\langle t_{\rm cc} \rangle$ [Myr] & $t_{\rm cc}[t_{\rm rlx}]$ \\
\hline
6kw6Scalo3 & $6122$ & $11$ &  $5.6$ & $1.0$ & $0.18 \pm 0.05$ \\
12kw6Scalo3 & $12288$ & $7$ & $13.8$  & $2.4$  & $0.17 \pm 0.06$ \\ 
\hline
\end{tabular}
\note{The first column shows the name of the model that describes the basic features of the model. The first numbers together with \textit{k} indicate the number of stars in the model. The number following w is the $W_0$ parameter of the initial King density distribution. The next set of letters is the name of the initial mass function (IMF) and the last number is the power of 10 of the upper-to-lower-mass-ratio of the IMF. For instance 6kw6Scalo3 is the code name of a system of $6k=6\times1024=6122$ stars distributed according to a King density profile with parameter $W_0=6$ and masses given by a Scalo IMF with $m_{\rm upper}/m_{\rm lower} = 10^3$. For all models presented here, the upper mass limit is 100 $M_\odot$ and the lower mass limit is 0.1 $M_\odot$. The second column shows the number of stars in the simulation. The third indicates the number of runs for the model. The fourth column gives the average half mass relaxation time in Myr. The fifth column shows the average core collapse time in Myr calculated using \texttt{Myriad}. In the sixth column, the calculated core collapse time in units of $t_{\rm rlx}$ is presented.}
\label{table1}
\vspace{1cm}
\end{table*}

\begin{figure*}[!ht]
\begin{minipage}[t]{0.5\linewidth}
\centering
\includegraphics[height=8cm, width=1\linewidth]{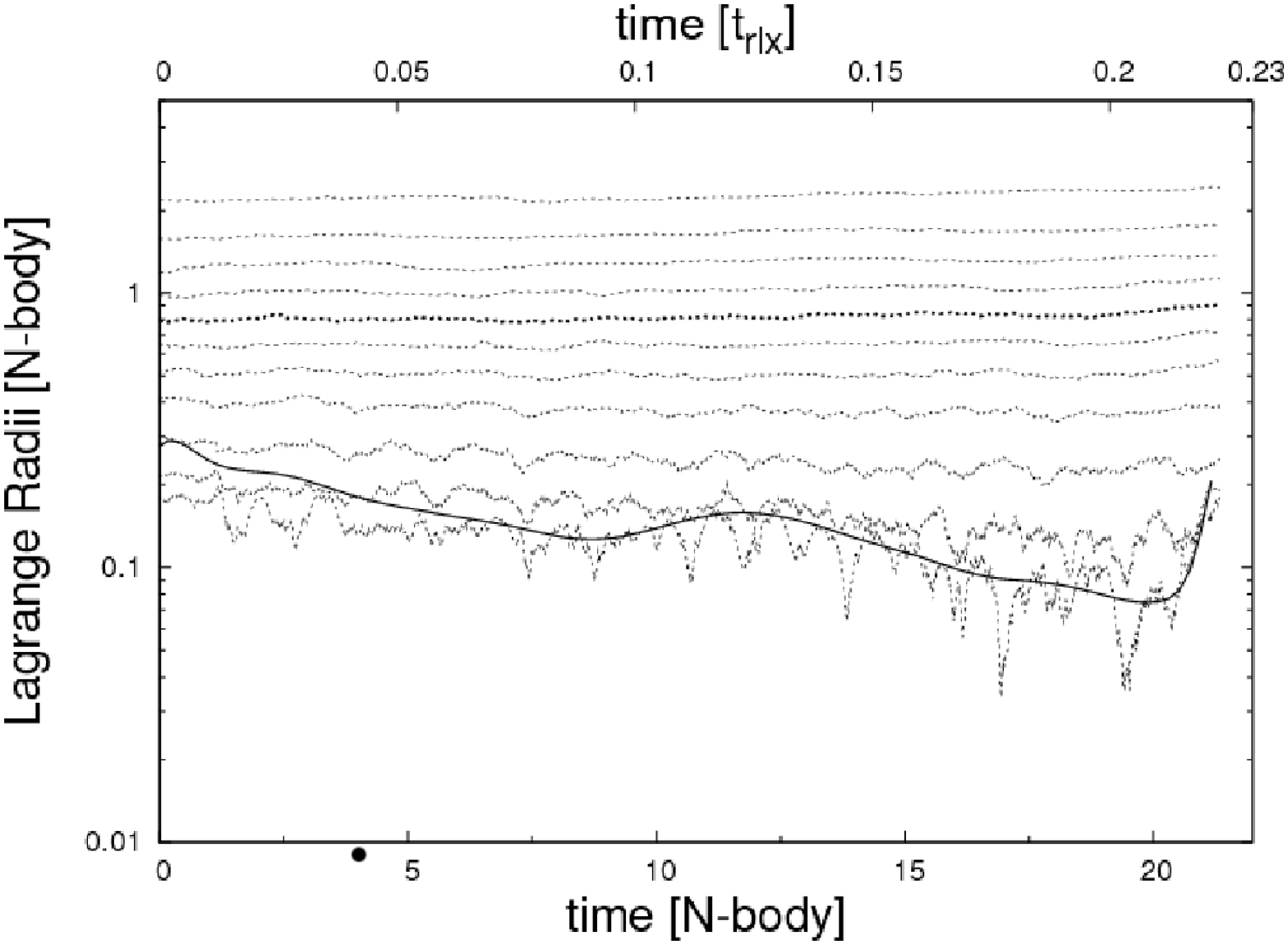}
\caption{Core radius (heavy black line) and Lagrangian radii containing $90\%$, $80\%$,  $70\%$, $60\%$, $50\%$, $40\%$, $30\%$, $20\%$, $10\%$, $5\%$, and $3\%$ (dashed lines from top to bottom) of the total mass. The half mass radius is indicated by a heavier dashed line. The initial condition was a King model with $W_0 =6$ containing 6144 stars. A Scalo mass function with mass limits $m_{\rm lower} =  0.1 M_\odot$ and $m_{\rm upper} = 100 M_\odot$ was chosen as the initial mass function. Core collapse is reached at $t_{\rm cc}\simeq 20.6$ N-body time units or $t_{\rm cc} \simeq 0.21 t_{\rm rlx}$, which is the time of formation of the first hard binary.}
\label{fig:fig6K}
\end{minipage}
\hspace{0.5cm}
\begin{minipage}[t]{0.5\linewidth}
\centering
\includegraphics[height=8cm, width=1\linewidth]{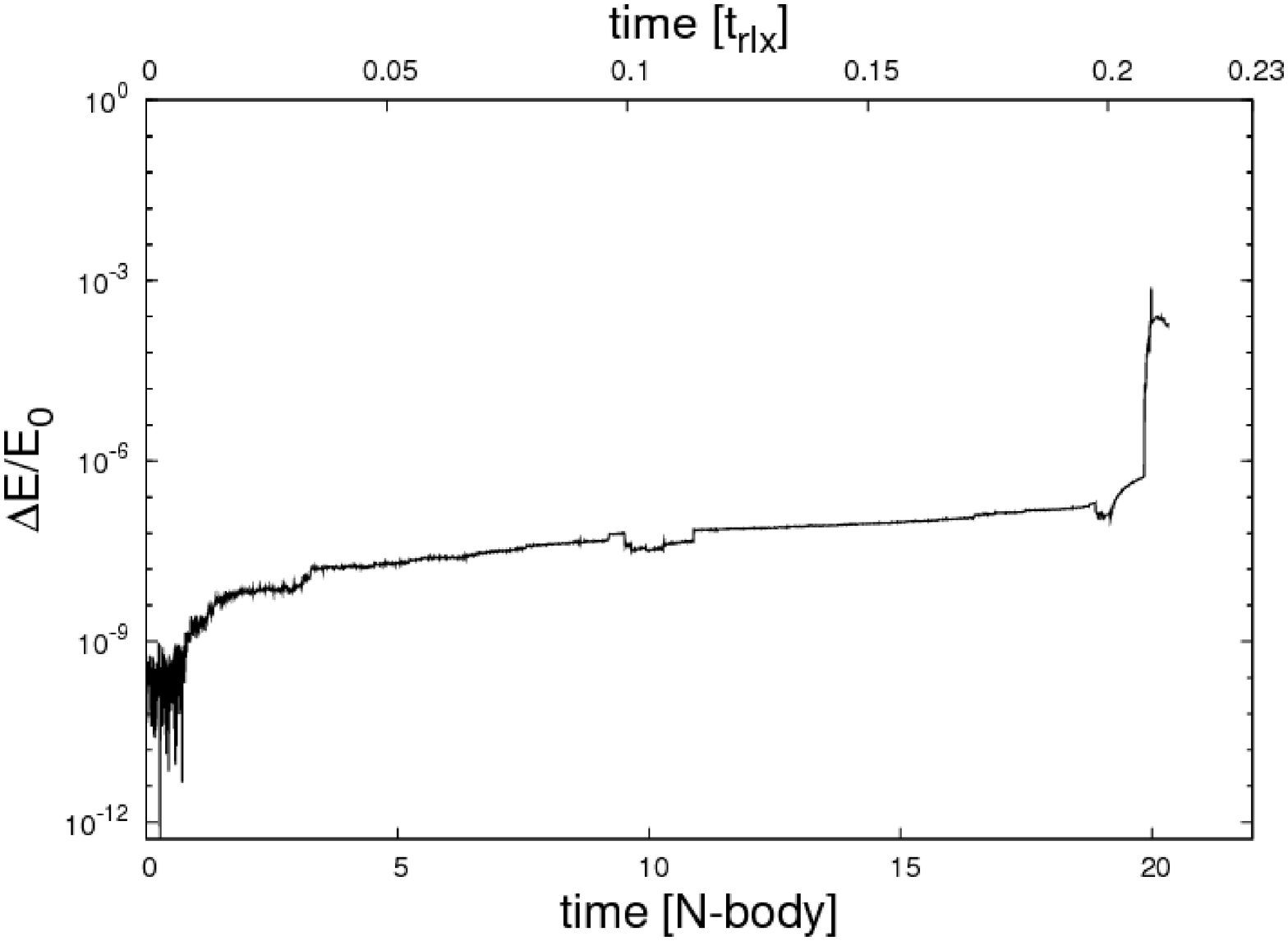}
\caption{Cumulative error in the energy ($\Delta E/E_{0} = (E_{\rm t} - E_{0})/E_{0}$) as it evolves in time for the simulation of Figure \ref{fig:fig6K}.}
\label{fig:fig6Ka}
\end{minipage}

\end{figure*}

\begin{figure*}[!ht]
\begin{minipage}[t]{0.5\linewidth}
\centering
\includegraphics[height=8cm, width=1\linewidth]{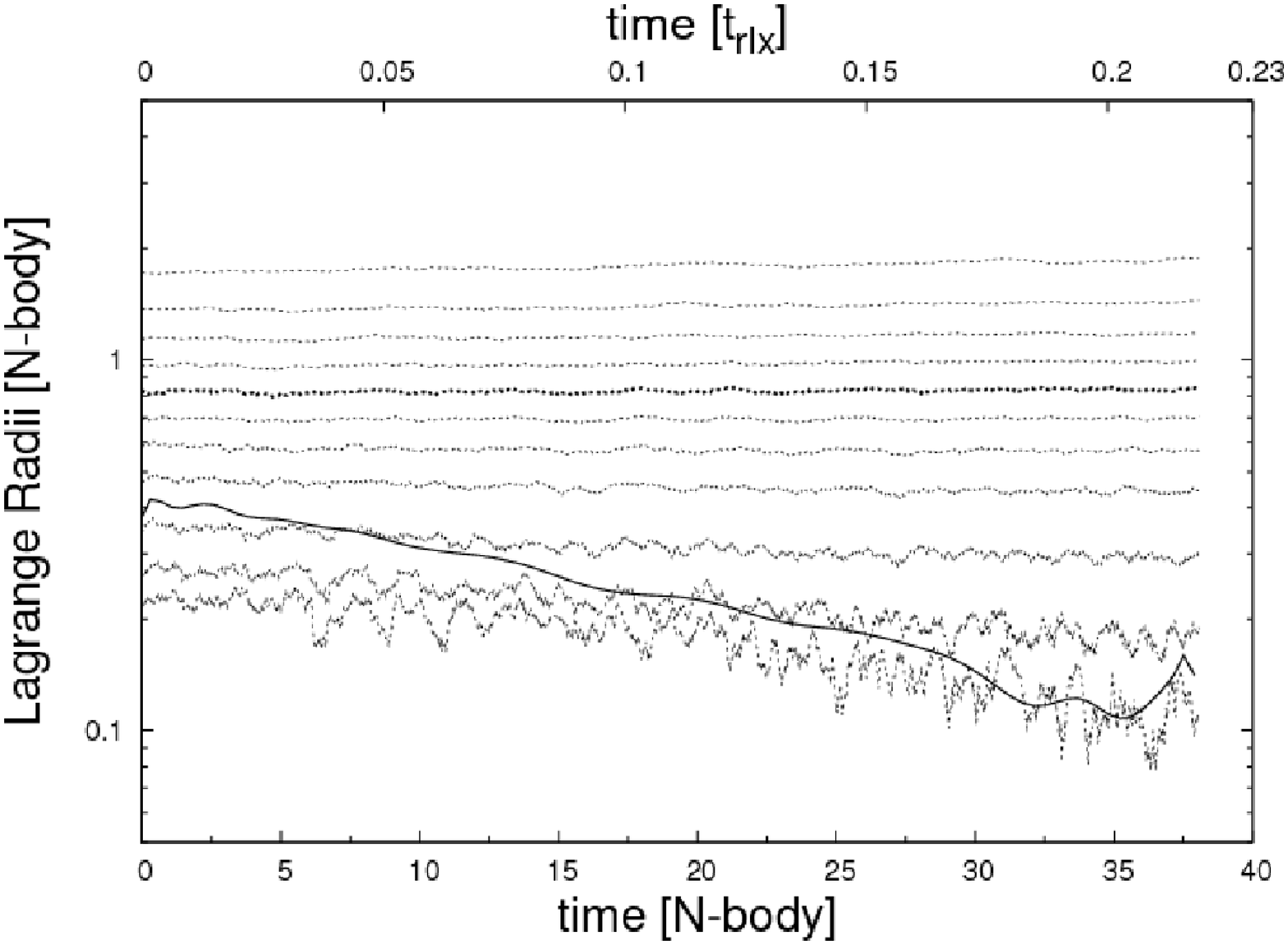}
\caption{Core radius (heavy black line) and Lagrangian radii containing $90\%$, $80\%$,  $70\%$, $60\%$, $50\%$, $40\%$, $30\%$, $20\%$, $10\%$, $5\%$, and $3\%$ (dashed lines from top to bottom) of the total mass. The half mass radius is indicated by a heavier dashed line. The initial condition was a King model with $W_0 =6$ containing 12288 stars. A Scalo mass function with mass limits $m_{\rm lower} =  0.1 M_\odot$ and $m_{\rm upper} = 100 M_\odot$ was chosen as the initial mass function. Core collapse is reached at $t_{\rm cc}\simeq 37.2$ N-body time units or $t_{\rm cc} \simeq 0.21 t_{\rm rlx}$, which is the time of formation of the first hard binary.}
\label{fig:fig12K}
\end{minipage}
\hspace{0.5cm}
\begin{minipage}[t]{0.5\linewidth}
\centering
\includegraphics[height=8cm, width=1\linewidth]{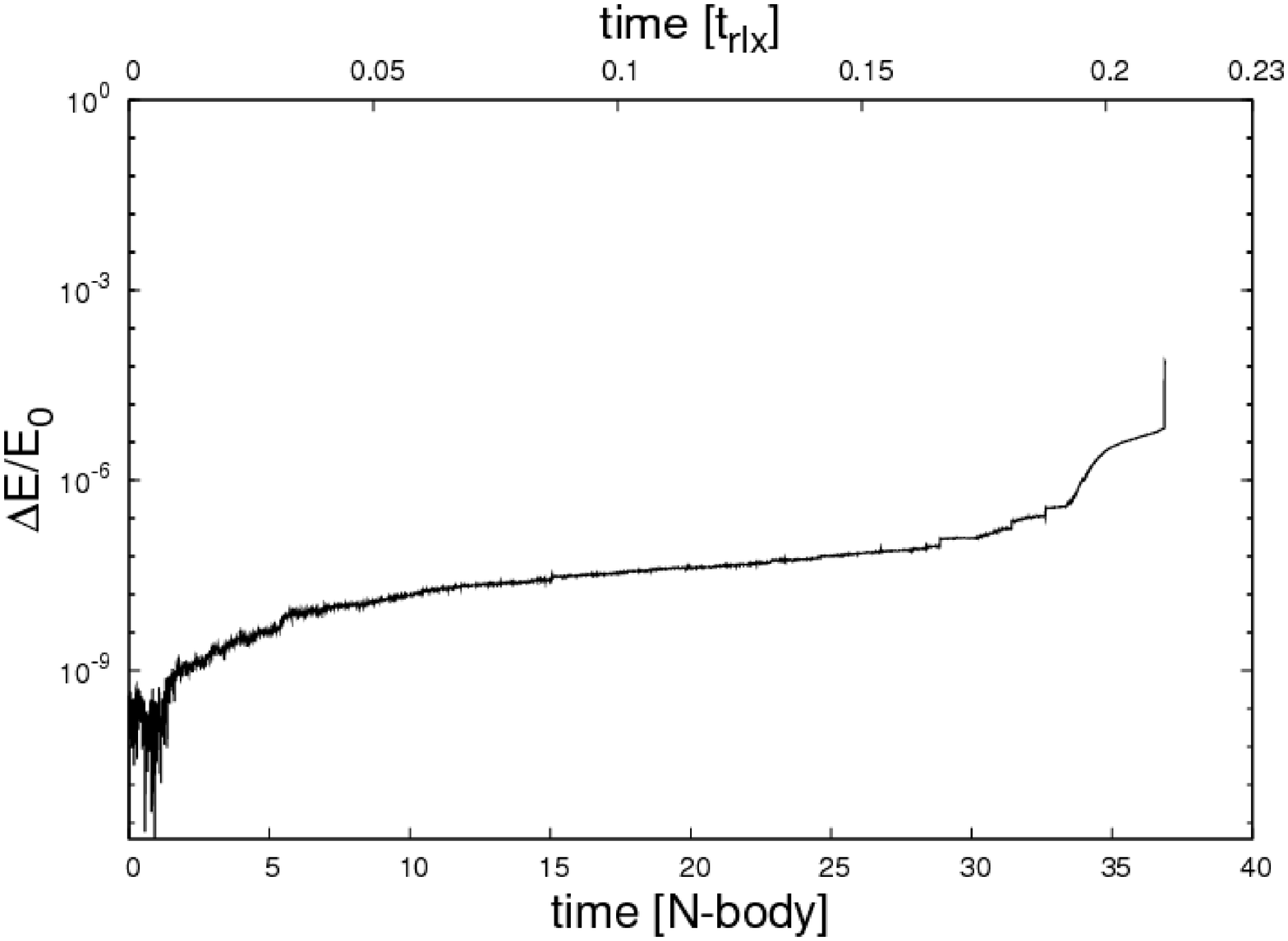}
\caption{Cumulative error in the energy ($\Delta E/E_{0} = (E_{\rm t} - E_{0})/E_{0}$) as it evolves in time for the simulation of Figure \ref{fig:fig12K}.}
\label{fig:fig12Ka}
\end{minipage}

\end{figure*}

We performed simulations of star clusters with initial density distribution of a $W_0 = 6$ King model \citep{King} up to core collapse. We used a Scalo \citep{Scalo} initial mass function (IMF) with lower and upper limits of $0.1 M_{\odot}$ and $100 M_{\odot}$, respectively. We varied the number of stars from $N = 6122$ to $N = 12\,288$ and found the core collapse time $t_{\rm cc}$ as a fraction of the half mass relaxation time $t_{\rm rlx}$ of the systems. The systems chosen are similar to those used in \cite{Zwart2003}, where the simulations were performed using the \texttt{Starlab} package including stellar mass loss from stellar evolution. The results are presented in Table 1. As the core collapse time we consider the time of the creation of the first dynamically formed hard binary system in agreement with \cite{Zwart2003}. We performed 18 simulations and our result for the core collapse time is $t_{\rm cc} \simeq (0.17\pm 0.05) t_{\rm{rlx}}$ . The result is slightly smaller than \cite{Zwart2003}, which is $t_{\rm cc} \simeq (0.19 \pm 0.08) t_{\rm{rlx}}$ . This difference in results may be due to the lack of stellar mass loss and stellar evolution in \texttt{Myriad} which, according to \cite{Zwart2003}, tends to delay core collapse. 

The evolution of Lagrange radii and core radius in one of the simulations containing $N = 6122$ stars is shown in Figure \ref{fig:fig6K}. For a clearer visual result, we have applied a cubic spline interpolation to the core radius. The cumulative error in the energy for the same simulation is presented in Figure \ref{fig:fig6Ka}. In Figs. \ref{fig:fig12K} and \ref{fig:fig12Ka}, we present the same results for one of the simulations containing $N = 12\,288$ stars. The cumulative energy error in all simulations remained below $ 10^{-3}$.

\section{Discussion}

\paragraph{Sources of energy errors}

There are unavoidable numerical errors in all N-body simulations. Here we discuss the source of numerical errors in simulations with \texttt{Myriad} and also how the total CPU time is divided between the different parts of the code.

The pre-collapse error does not depend on the binary evolution since only some short-lived close encounters and soft binaries need the special treatment of the binary module of the code. This error depends mainly on the accuracy parameter $\eta$ as described in Sect. 3.1. The post-collapse error is more complex to analyze, because long-living hard binaries form and interact with individual particles and other binaries. In some cases, multiple subsystems form. All of these uncertainties need to be treated very carefully. Since the error in the evolution of isolated hard binaries and multiples can be very small as described in Sect. 3.3, the relatively large errors in the energy after core collapse (see Figures \ref{fig:fig12}, \ref{fig:fig12a}, \ref{fig:fig14}, \ref{fig:fig6Ka}, and \ref{fig:fig12Ka}) come mainly from the interactions of binaries and multiples with the rest of the system. The external perturbations in a binary are controlled by the dimensionless parameter $\gamma_{\rm pert}$ discussed in Sect. 2.2.3. This parameter plays role in the total energy error, since if it is large enough, some of the perturbations are ignored and systematic errors are introduced into the binary evolution and the evolution of stars that lie close to the binary. Our choice of $\gamma_{\rm pert}$ is $10^{-7}$, which probably needs adjustment to systems containing stars with large mass ratios.

When a binary or multiple sub-system forms according to the rules presented in Sect. 2.2.3, the transformation of coordinates from the center of mass of the cluster to the local center of mass of the binary or multiple, introduce another source of energy error. If that binary is a soft binary, it is usually required to form close to periastron passage and is deformed close to apastron passage once every period, until it transforms into a hard binary or its eccentricity becomes more than $1$. This continuous transformation of coordinates introduces systematic errors into the simulation. Systematic errors are also introduced when a third star orbits with high eccentricity around the center of mass of a hard binary. According to the rules of \texttt{Myriad}, the triple system would have to form and deform often depending on the hardness of the hard binary and the eccentricity of the third star. In this case, another source of error may introduce large errors into the simulation: the third star would have to jump from small time steps, when it is a member of the triple, to large time steps, when it is considered to be an escaper from the triple and evolves as a single star in the whole system. 

Finally, it is not clear what the optimal parameters are for the construction and destruction of a binary or a multiple system, because there are many different cases. A multiple sub-system for example may have formed around a soft binary or hard binary or even a close encounter between two stars may occur, and a particle can escape the sub-system under different  conditions. The parameters used in \texttt{Myriad} are those described in Sect. 2.2.3, but they need further examination to reduce systematic energy errors. For example, it is unclear when an interaction between 2 hard binaries can lead to the formation of a multiple sub-system with 4 stars and if such a multiple is formed, it is not easy to determine under the conditions under which it would be destroyed forming smaller sub-systems.

From the above, it is clear that the main source of error in the post-collapse part of a simulation, where hard binaries form in a dense environment, depends on the treatment of the interactions between the binaries and their surrounding environment and on the choices we make for the rules of creation and dissipation of multiples. The sudden jumps in the time step of particles that interact with hard binaries are another important source of error. Finally, the dense environment itself introduces large errors, because there are relatively many stars sharing small time steps and more time steps per time unit, introduce large errors in the energy.

\paragraph{CPU time}

The total CPU time $T_{\rm CPU}$ taken for a simulation with \texttt{Myriad} can be divided into the CPU time needed for the different modules of the code
\begin{equation}
 T_{\rm \textsc{cpu}} = T_{\rm \textsc{bin}} + T_{\rm \textsc{herm}} + T_{\rm \textsc{force}} + T_{\rm \textsc{check}} + T_{\rm \textsc{diag}},
\end{equation}

where $T_{\rm \textsc{herm}}$ is the CPU time for the predictor, corrector, and finding the time steps of the Hermite 4 scheme. In the simulation of an $1K$ system up to and beyond core collapse presented in Sect. 3.4.1, this time is less than $1\%$ of the total CPU time.

The parameter $T_{\rm \textsc{force}}$ is the CPU time required by GRAPE-6 to calculate the accelerations and identify the neighbors of particles. We reduced this CPU time by asking GRAPE-6 to search for and return neighbors only for particles that have small time steps. We restricted the search for neighbors to the particles that belong to the first 3 time blocks, without any change in the accuracy of the code. For the simulation of Sect. 3.4.1, $T_{\rm \textsc{force}} \simeq 51 \% T_{\rm \textsc{cpu}}$. As the number of particles in a simulation increases, $T_{\rm HERM}$ is expected to increase according to the rule $N\log{N}$.

The parameter $T_{\rm \textsc{bin}}$ is the CPU time spent in the evolution of binary and multiple systems that form in a cluster. It depends on the number of binaries formed, on the time step used for their evolution, and on the number of stars that are considered as perturbers of a binary system. If primordial binaries exist, this CPU time is expected to decelerate the simulation considerably. In the simulation of Sect. 3.4.1 where no primordial binaries exist, and hard binaries formed just before core collapse, $T_{\rm \textsc{bin}} \simeq 42\% T_{\rm \textsc{cpu}}$. In the same simulation, $T_{\rm \textsc{bin}}$  before core collapse is insignificant.

$T_{\rm \textsc{check}}$ is the CPU time needed to check for encounters and binaries in the system. We limited this check to particles that have small time steps, since the time step given by Eq. (\ref{Aarseth}) is an indicator of the dynamics around a particle. If at a given time the time step is small, that means that there is a probability that the particle is about to undergo a close encounter with some other particle, and a search for neighbors is required. If the time-step is large, then the particle probably lies in a relatively sparce environment and a close encounter with other particles is improbable, at least at the current time. $T_{\rm \textsc{check}}$ depends on the density of stars at the cluster center and for this reason grows for clusters that are close to core collapse. In the simulation described in Sect. 3.4.1, we found $T_{\rm \textsc{check}} \simeq 4\% T_{\rm \textsc{cpu}}$.

Finally, $T_{\rm \textsc{diag}}$ is the CPU time required to calculate the global parameters of the cluster, such as the core radius, remove escaping stars, and write output snapshots to a file. The most ``expensive'' in terms of CPU time is the calculation of the core radius of the system, because it evaluates the density around each particle. GRAPE-6 is used for this calculation, but the process is still relatively slow. For the simulation of Sect. 3.4.1, it is $T_{\rm \textsc{diag}} \simeq 2 \%T{\rm \textsc{cpu}}$.

\section{Conclusions}

We have developed a new C++ N-body code called \texttt{Myriad}. The code can simulate the evolution of star clusters with excellent accuracy, while its computational speed is satisfactory. The accumulated relative error in the total energy in the simulation of $N=16\,384$ stars with a Salpeter IMF ($m_{\rm upper} = 5 M_{\odot}$, $m_{\rm lower} = 0.5 M_{\odot}$), evolved up to and beyond core collapse, is smaller than $2.5 \times 10^{-5}$, while the error in the energy per quarter of N-body time unit is smaller than $6\times10^{-6}$. The computational time needed for this simulation was less than $3$ days, which is comparable to the computational time of most N-body codes, but could be improved. The results for the core collapse time of star clusters with an IMF or with equal-mass stars are also comparable with those found in the literature or produced by other codes such as \texttt{Starlab}. In \citet{StarNbodyCompare}, \texttt{Starlab} and \texttt{NBODY4} are compared in detail, and \texttt{Myriad}  successfully reproduces the result for the core collapse time of an equal-mass Plummer model consisting of $1K$ particles. 

We now summarize the most important innovations introduced by \texttt{Myriad}:
\begin{enumerate}
 \item The accuracy parameter $\eta_{\rm b}$ for the evolution of binary and multiple sub-systems is adjusted for every one of them by the time of its formation, while in other codes it is fixed to some value. Variable $\eta_{\rm b}$ makes the transition of time steps between the H4 scheme and the binary module smooth. In the future, $\eta_{\rm b}$ may change during the binary evolution to retain the accuracy in desirable limits.
 \item The algorithms of \citet{time_symmetric1} and \citet{tsHermite4} have been successfully applied to the $P(EC)^3$ scheme for the evolution of binary and multiple sub-systems. The error in the energy of these systems does not increase linearly with time, but remains bounded within some limits in most of the cases.
 \item We have introduced new rules for creating triple systems when single stars are approaching binaries. We do not use any distance criterion, but the dimensionless perturbation parameter $\gamma$ defined in Eq. (\ref{gamma}), which depends on the hardness of the binary, the distance of the third star and their masses. We use the same parameter to remove a star from a multiple sub-system.
\end{enumerate}

There are 4 important parameters of \texttt{Myriad} that balance between speed and accuracy. These are:
\begin{enumerate}
 \item The accuracy parameter $\eta$, used for Aarseth's time step criterion given in Eq. (\ref{Aarseth}). This is set by the user before beginning a simulation with \texttt{Myriad}. A typical value of $\eta$ is 0.01.
 \item The accuracy parameter $\eta_{\rm b}$ for the time step criterion in Eq. (\ref{dt_b}), used for the evolution of binary and multiple subsystems. This parameter is adjusted during runtime until a smooth transition between the time step of the binary and the Hermite integrator is achieved.
 \item The critical value of the perturbation parameter $\gamma_{\rm crit}$, discussed in Sect. 2.2.3, above which a particle that perturbs a binary system, becomes a member of the same binary, forming a triple system. The value of this parameter is set in \texttt{Myriad} to be $0.125$ for soft binaries and $0.015625$ for hard binaries.
 \item The critical value of the perturbation parameter $\gamma_{\rm pert}$ above which a particle is considered as a perturber of a binary or multiple subsystem (see Sect. 2.2.3: Perturbers). A typical value of this parameter that is used throughout the present work is $10^{-7}$, but it is automatically increased if it results in a large number of perturbers per binary. The maximum number of perturbers per binary system is set to $100$.
\end{enumerate}

From the results of the present work, we conclude that \texttt{Myriad} has the basic structure of a code that could be used to model the evolution of star clusters with IMBHs at their centers and study the effects of the presence of an IMBH on the system and the possibility of its ejection due to encounters and collisions with other stellar-mass black holes.
To achieve this, additional improvements to the code will be required so that more physical phenomena are included in the simulations. One of the improvements is to include post-Newtonian dynamics into the equations for the evolution of close binary black holes. Because of the modular structure of the code, this improvement could be made by simply adding a function under the \textsc{binary} class. This function would calculate the appropriate post-Newtonian terms (up to $2.5$ \textit{PN}) for the acceleration and its first derivative, following  the formalisms found in \citet{PostNewtonian2}, \citet{PostNewtonian3}, and \citet{PostNewtonian1}. Another extension that would make the \textsc{binary} class more complete in treating binary black holes is the inclusion of the relativistic effect of the asymmetric emission of gravitational radiation and recoil velocity, during the merging of two orbiting black holes. According to numerical relativity results, when two spinning black holes collide, gravitational radiation could be emitted asymmetrically. This would lead to a recoil velocity in the resulting black hole that might be as high as $4000 km/s$ \citep{Recoil0,Recoil4, Recoil5,Recoil7,Recoil8}, depending on the mass ratio of the initial black holes and the directions of their spins, but this velocity might be significantly suppressed by the relativistic alignment of the spins \citep{Recoil9}.  Tt is not easy, of course to include numerical relativity in an N-body code, but semi-analytical formulae, coming from fitting between numerical relativity results and post-Newtonian theory \citep{Recoil1, Recoil2, Recoil3} to determine the direction and magnitude of the recoil velocity. These semi-analytical formulas could be easily included in the binary module of \texttt{Myriad}, making the code a tool capable of reproducing collisions of black holes realistically. Those two extensions include effects beyond the Newtonian theory of gravitation, and with extensive tests and comparisons to theoretical and already published results, will be presented in a future paper. After some development, \texttt{Myriad} should be able to study the behavior of BHs that come from mergers in a cluster of stars, the possible changes in the structure of the cluster, and the possibility of the escape of IMBHs from clusters \citep{Recoil6}, after collisions with stellar-mass BHs.

Additional improvements to the code could be made by adding stellar and binary evolution and the influence of the Galactic tidal field. This would make the simulation more realistic, because mass loss from stellar evolution and the tidal force of the Galaxy play an important role in the evolution of the whole cluster. Because of the code structure, the addition of stellar evolution may be achieved by simply including the appropriate function as part of the \textsc{particle} class, while binary evolution could be included as a simple function under the \textsc{binary} class. 

Finally, improvements could be made to the speed of the code, by connecting the \textsc{particle} classes in a binary search tree. The search for close encounters and perturbers would then become faster and more accurate, since GRAPE-6 memory overflow would be avoided. Another change in the structure of the code would be to divide the calculations between the CPU, GRAPE-6, and the GPU, such that all three parts of the hardware cooperate in making the code even faster. A version of the code that uses the GPUs instead of GRAPE-6 to calculate the forces, is under development. Until now, this version has usee the Sapporo library \citep{sapporo} to emulate the GRAPE-6 functions on the GPUs and a shared time step for all particles. Results of preliminary tests show that, in agreement with the introductory paper of Sapporo, the speed of the code using 4 Nvidia Tesla C1060 GPUs is comparable with the speed of the code when GRAPE-6 is used. Parallelization of the code, so that it could run on clusters of computers attached to accelerating hardware of the GRAPE-6 family or GPUs, is an improvement that needs to be made to be able to simulate realistically large star-systems.

\appendix

\section{Units and basic equations}

\paragraph{N-body Units:}
The units adopted in the simulations are the usual N-body units\footnote{\url{http://en.wikipedia.org/wiki/Natural_units#N-body_units}} \citep{NbodyUnits}, where $G = M_{t} = R_{\rm V} = 1$, and $G$ is the gravitational constant, $M_{\rm t}$ is the total mass of the system, and $R_{\rm V}$ is its virial radius. In these units, the total energy is $E_{\rm t} = -1/4$. Transformation to physical units can be made if the total mass $m_{\rm t}$ and the virial radius $r_{\rm v}$ are known in physical units. If the mass $m_{\rm t}$ is given in solar masses [$M_\odot$] and $r_{\rm v}$ in parsecs [$pc$], the mass $m_i$ of a star $i$ in solar masses
\begin{equation}
 m_i = M_i m_{\rm t}  \; \; [M_\odot].
\end{equation}
Any distance $R$ in N-body units is transformed to [pc] using
\begin{equation}
 r =  R r_{\rm v} \; \; [pc].
\end{equation}
In addition, there are functions for transforming time and velocity from N-body ($T,V$) to physical units ($t,v$)
\begin{equation}
 t = T T^*  \; \; [Myr]
\end{equation}
and
\begin{equation}
 v = V  V^*  \; \; [kms^{-1}],
\end{equation}
where 
\begin{equation}
V^* = 6.557 \times 10^{-2} \Big( \frac{m_{\rm t}}{r_{\rm v}} \Big)^{1/2} \; \; [kms^{-1}]
\end{equation}
and 
\begin{equation}
T^* = 14.94 \Big( \frac{r_{\rm v}^3}{m_{\rm t}} \Big)^{1/2} \; \; [Myr].
\end{equation}
The numerical factors $6.557 \times 10^{-2}$ and $14.94$ come from expressing $1 \times 10^{-5} (G M_\odot/L^*)$ and $({L^*}^3/GM_\odot)^{1/2}$ in cgs units, respectively, where $L^*$ is taken to equal $1 \; pc$ and $G$ and $M_\odot$ are expressed in cgs units.

In the following, we follow the definitions of \citet{AarsethBook}. 
\paragraph{Half-mass radius:}
The half mass radius $r_{\rm h}$ is defined as the radius of the sphere that has its center at the center of mass of the star cluster and contains half of the mass of the total system.
\paragraph{Half-mass relaxation time:}
The equation for the half-mass relaxation time is adopted from \citet{AarsethBook} and \citet{Spitzer}
\begin{equation} \label{relax}
 t_{\rm rlx} = 0.138 \Big( \frac{N r_{\rm h}^3 }{G\bar{m}} \Big)^{1/2} \frac{1}{\ln{\Lambda}},
\end{equation}
where $\bar{m}$ is the mean mass and $\ln{\Lambda}$ is the Coulomb logarithm \citep{Spitzer} given by
\begin{equation}
 \ln{\Lambda} = \ln{\gamma N},
\end{equation}
where $N$ is the number of stars and the factor $\gamma$ is usually chosen to be $0.4$ \citep{AarsethBook}, but in some cases values of around $0.1$ may be more appropriate \citep{Giersz}. A typical value of $\ln{\Lambda}$ used in N-body simulations is 10.
\paragraph{Crossing time:}
The time to cross the cluster is given by 
\begin{equation} \label{crossing}
 t_{\rm cr} \simeq 2 \sqrt{2} \Big( \frac{r_{\rm v}^3}{GN\bar{m}} \Big)^{1/2},
\end{equation}
where $\bar{m}$ is the mean mass of the system. In N-body units, it is $T_{\rm cr} = 2\sqrt{2}$. Equation (\ref{crossing}) is derived from 
\begin{equation}
 t_{\rm cr} = 2 r_{\rm v}/\sigma,
\end{equation}
where $\sigma$ is the $rms$ velocity dispersion given at equilibrium by
\begin{equation}
 \sigma \simeq \sqrt{\frac{G N\bar{m}}{2 r_{\rm v}}}.
\end{equation}

\paragraph{Core radius:}
The core radius is defined by the equation
\begin{equation} \label{rcc}
 r_{\rm c} = \sqrt{\frac{\sum_{i} \rho_i^2 | \textbf{r}_i - \textbf{r}_{\rm d}|^2}{\sum_i \rho_i^2}},
\end{equation}
where $\rho_i$ is the local density around star $i$ given by
\begin{equation} 
\rho_i = \frac{\sum_{j=1}^{5} m_j} {\frac{4}{3}\pi r_{\rm 6}^3}
\end{equation}
and $r_{\rm 6}$ is the distance to the sixth nearest neighbor of star $i$, while 
$r_{\rm d}$ is the position of the density center whose definition follows.
\paragraph{Density center:}
The density center of the star cluster is defined to be
\begin{equation} \label{density}
 \textbf{r}_{\rm d} = \frac{ \sum_{i} \rho_i \textbf{r}_i}{\sum_i \rho_i}.
\end{equation}

\paragraph{Core collapse time:}
For an equal-mass system, the time for core collapse is \citep{Spitzer}
\begin{equation} \label{cc_1}
 t_{\rm cc} \simeq 15 t_{\rm rlx}.
\end{equation}
For a system containing stars with different masses distributed according to a King density profile \citep{King}, the core collapse time becomes \citep{Zwart2003}
\begin{equation}\label{cc_2}
 t_{cc} \simeq 0.2 t_{\rm rlx}.
\end{equation}

In an N-body simulation, a clearer definition of the initial core collapse time is sometimes required since in some cases it is unclear when exactly the core shrinks to a minimum size and then starts its expansion. In many publications and for more accurate comparison between different codes \citep{StarNbodyCompare}, the core collapse time is considered to be the time of the formation of the first long-living hard binary with binding energy greater than $100kT$, where $kT$ is the equivalent to the thermal energy of a gas given for a cluster of $N$ stars by
\begin{equation}
 E_{kin} = \frac{3}{2} N k T
\end{equation}
and $E_{kin}$ is the total kinetic energy of the cluster.
In other publications \citep{Primordial1}, linear fitting is applied to the diagram of $r_{\rm c}$ versus $\rm{time}$ and the core collapse time is determined by the intersection of two lines, one being the fitting line of data before the core reaches its first minimum and the other being the fitting line after the core reaches its minimum. 

In some other publications, the core collapse time is simply found by observing the diagram $r_{\rm c}$ versus $\rm{time}$ where there is a clear and sharp minimum.

In this work, the time of core collapse is determined both by simply observing the diagram of the evolution of core radius with time and by using the formation of the first hard binary with binding energy greater than the $100kT$ criterion.

\begin{acknowledgements}
We would like to acknowledge the authors of \texttt{Starlab} for making their code freely available, and the creators of the \textit{Art of Computational Science} website for their freely available codes and instructive books. We also would like to thank Kleomenis Tsiganis, for his very helpful suggestions and the anonymous referee for his/her instructions that helped us improve and correct the manuscript. This work was supported by the German Science Foundation (DFG) via SFB/TR7 on ``Gravitational Waves'' and by the Greek GSRT Pythagoras I program. The use of the GRAPE-6 system was supported by a grant by the Empirikion Foundation. SK would like to thank the Eberhard-Karls University of T\"{u}bingen, where part of the code was written, for hospitality and support.
\end{acknowledgements}

\bibliographystyle{aa}
\bibliography{refs}		

\end{document}